\documentclass[journal]{IEEEtran}

\usepackage{cite}
\usepackage{amsmath,amssymb,amsfonts}
\usepackage{algorithmic}
\usepackage{placeins}
\ifCLASSINFOpdf

\usepackage[export]{adjustbox}  
\usepackage{lipsum,booktabs}
\usepackage{verbatim} 

\usepackage{graphicx}
\ifCLASSOPTIONcompsoc
    \usepackage[caption=false, font=normalsize, labelfont=sf, textfont=sf]{subfig}
\else
\usepackage[caption=false, font=footnotesize]{subfig}
\fi

\usepackage{ragged2e}
\usepackage{subfig}
\usepackage{multirow}

\makeatletter                        
\def\Cline#1#2{\@Cline#1#2\@nil}
\def\@Cline#1-#2#3\@nil{%
  \omit
  \@multicnt#1%
  \advance\@multispan\m@ne
  \ifnum\@multicnt=\@ne\@firstofone{&\omit}\fi
  \@multicnt#2%
  \advance\@multicnt-#1%
  \advance\@multispan\@ne
  \leaders\hrule\@height#3\hfill
  \cr}
\makeatother

\usepackage{algorithm}
\usepackage{array}
\usepackage{textcomp}
\usepackage{stfloats}
\usepackage{url}
\begin{document}

\title{Implementation and Evaluation of \\SiPM-based Photon Counting Receiver \\for IoT Applications}
\author{Yangchun Li, and Danial Chitnis
\thanks{Manuscript received XX, XX, XXXX; revised XX, XX, XXXX; accepted XX, XX, XXXX. Date of publication XX, XX, XXXX; date of current version XXXX, XX, XXXX. \textit{(Corresponding author: Danial Chitnis.)}}  
\thanks{Yangchun Li is with the School of Engineering, Institute for Integrated Micro and Nano Systems, University of Edinburgh, EH9 3FF Edinburgh, U.K. (e-mail: y.li-257@sms.ed.ac.uk).}
\thanks{Danial Chitnis is with the School of Engineering, Institute for Integrated Micro and Nano Systems, University of Edinburgh, EH9 3FF Edinburgh, U.K. (e-mail: d.chitnis@ed.ac.uk).}
\thanks{Digital Object Identiﬁer XXXXXXXX}}

\markboth{IEEE INTERNET OF THINGS JOURNAL,~Vol.~XX, No.~X, ~XXXX, xx xx 2024}%
{Shell \MakeLowercase{\textit{et al.}}: A Sample Article Using IEEEtran.cls for IEEE Journals}

\IEEEpubid{\begin{minipage}{\textwidth}\ \centering
Copyright (c) 2024 IEEE. Personal use of this material is permitted. \\ However, permission to use this material for any other purposes must be obtained from the IEEE by sending a request to pubs-permissions@ieee.org.
\end{minipage}}


\maketitle

\begin{abstract}

Silicon Photomultipliers (SiPMs) are photon-counting detectors with great potential to improve the sensitivity of optical receivers. Recent studies of SiPMs in communication focus on the speed rather than the power consumption of the receiver. The gain bandwidth product (GBP) of the amplifiers in these post-SiPM readout circuits is significantly higher than the target data rate. Additionally, the SiPM experiments for optical communication are performed using an offline method which uses instruments including oscilloscopes and personal computers to process chunks of the transmitted data. In this work, we have developed an embedded real-time ﬁeld-programmable gate array (FPGA) based system to evaluate a commercially available 1~$mm^2$ SiPM. The implemented real-time system achieves data rates from 10 kbps to 1 Mbps with a bit error rate (BER) below $10^{-3}$ approaching the Poisson limit. Results showed that reducing either the dark count rate or increasing the data rate leads to lower dark counts per bit time, hence less power penalty to maintain a probability of error (PE) of $10^{-3}$. The numerically simulated results indicated that to maintain the Poisson limit, the minimum GBP of the amplifier in the post-SiPM readout circuit is 120~MHz based on the proposed setup within the tested data rates. This GBP limitation is determined by the noise floor of the read-out circuit. The analysis of the minimum GBP and electrical power consumption of the receiver in photon counting and BER enables the potential future adoption of this receiver technology when high optical sensitivity is required, such as visible light communications (VLC) for low data rate Internet of Things (IoT) applications.

\end{abstract}


\IEEEpeerreviewmaketitle

\section{Introduction}

\IEEEPARstart{T}{he} demand for a wider communication spectrum has prompted the development of optical wireless communication, which could utilize the unlicensed spectrum in the communication link \cite{Haas1}. The ideal optical receiver for communication devices should have high sensitivity, low power consumption and cost, which could be reduced by integrating the electronic circuit in application specific integrated circuits (ASICs) or field programmable gate arrays (FPGAs). The most popular photon detectors in optical receivers are PIN photodiode \cite{Su}, \cite{Teli} and avalanche photodiode (APD) \cite{Nada}. Typically, the noise equivalent power (NEP) of a PIN photodiode is restricted to the other components in the receiver, including the trans-impedance amplifier (TIA). In contrast to the PIN photodiode, an APD provides an internal gain that is closer to the photodetection stage. However, the gain-dependent excess noise increases the NEP more than the PIN photodiode. A solution to minimize this noise is operating the APD in the photon-counting mode above its breakdown voltage, and alongside a quenching resistor as a single-photon avalanche diode (SPAD) \cite{Spinelli}, \cite{Cova}. However, the SPAD requires a recovery time, typically a few tens of nanoseconds, to quench the self-sustained avalanche current triggered by a photon detection event within its avalanche region of the p-n junction. Once a photon is absorbed within this region, the impact ionization will generate the avalanche photo-current, which leads to a voltage drop and subsequent quenching of the photo-current. The challenge created by the recovery times is the non-linear response of the output events when receiving a large number of photons in a short time. One method to reduce the recovery time is implementing an array of SPADs in parallel, known as silicon photomultipliers (SiPMs) or multiple pixel photon counters (MPPCs) \cite{Dolinsky}. This type of SPAD array has been explored in a wide range of applications, including time-of-flight positron emission tomography (TOF-PET) \cite{Akamatsu}, visible light communication (VLC) \cite{Long0}, \cite{Zubair1}, \cite{Zubair2}, \cite{Shenjie}, \cite{Kosman}, light fidelity (LiFi)  \cite{Haas2, Chen} and light detection and ranging (LIDAR) \cite{Runze, Zhao}. In recent years, complementary metal-oxide-semiconductor (CMOS) SPAD arrays for optical receivers were investigated \cite{Danial, Fisher}. Additionally, the advantages of passive quenched MPPC were demonstrated through theoretical analysis \cite{Guoqing}. Furthermore, research proved that using SiPM as a detector in the receiver will be more sensitive than the optoelectronic integrated circuit containing an APD at 1~Gbps with $10^{-3}$ bit error rate (BER)\cite{Zubair1}, \cite{Zimmermann}.

\IEEEpubidadjcol
Although recent developments in optical communication have prioritized enhancing the data rate, there is still a need to investigate the SiPM-based receiver in high sensitivity and low-speed performance, particularly for devices such as those used in the Internet of Things (IoT)\cite{Yosuf}. Recently, most of the experimental tests carried out on SiPM technology used in optical communication have employed offline data processing, which uses arbitrary waveform generators (AWGs) and high-bandwidth oscilloscopes to simulate a physical communication link. In the offline method, a pre-generated data set is transmitted via the link using an AWG, and the output of the SiPM is captured using the high bandwidth and sampling rate oscilloscope, which is later transferred to the personal computer (PC) for processing and calculating the BER. This offline method does not consider the effect of electronic circuits within a real-world transceiver. Also, the high bandwidth instruments used in SiPM experiments for optical communication are not realistic for evaluating real-world use cases for such devices, especially in terms of power consumption of the electronic circuitry. Furthermore, the offline data processing method involves a series of sequential steps, which can be time-consuming when attempting to achieve a BER smaller than $10^{-4}$ due to a large amount of data that needs to be captured and processed. In optical communication systems, a BER limit of 3.8 × $10^{-3}$ is commonly used to evaluate the performance since it is possible to enhance this limit to $10^{-15}$ by implementing a forward error correction (FEC) code with a 7\% overhead \cite{xiang}. However, some applications, including the body-area networks \cite{dokania} that prioritize low latency, high reliability, and low design complexity, may choose not to use FEC.

Implementing a photon-counting receiver based on the SiPM presents several challenges. Primarily, selecting suitable amplifiers must be undertaken at the early stages of implementation. To increase the SiPM's weak signal to a detectable level for subsequent processing, it is essential to determine the appropriate levels of gain and bandwidth that the post-readout circuit is working with. The relationship between the amplifier's gain and bandwidth is defined by the gain-bandwidth product (GBP), which is a parameter that characterizes the amplifier's capacity to deliver amplification across a spectrum of required frequencies. Previous studies \cite{Zubair1},\cite{Zubair2}, \cite{Shenjie}, \cite{Long2}, \cite{Jinjia} used large GBP amplifiers to increase the signal amplitude, sometimes without considering the minimum required GBP for the post-readout circuit. In addition, some systems may require a comparator based on specific modulation schemes, including on-off keying (OOK). For these systems, selecting the optimal threshold and hysteresis of the comparator is essential for signal noise mitigation and accurate digital pulse conversion.

Another challenge emerges when implementing the digital circuit for photon counting and accumulation. Unlike a typical circuit that can rely on synchronous clocks for receiving pulses and achieving continuous data reception, using high-frequency clock synchronization in a low data rate receiver is inefficient. Consequently, an alternative approach is needed to realize the real-time implementation of the SiPM-based photon-counting receiver.

In this paper, we investigated the feasibility of implementing a SiPM-based real-time receiver for IoT devices in optical communication. The effect of the electronic circuits, especially noise and digital circuits timing, was analyzed in system-level integration. The paper is organized as follows. Section II introduces the SiPM readout methods and previous results. Section III provides the SiPM operation background and characterization result. Section IV contains the experimental BER and power consumption results for the real-time SiPM-based optical receiver. The effect of the GBP in the receiver circuit and numerical simulation result was described in Section V. The result discussion and future works were provided in Section VI. Finally, section VII concludes this paper.

\section{SiPM Readout Methods and Previous Results}

In order to obtain the number of detected photons, a resistor is added to the anode or cathode of the SiPM to convert the total avalanche current into a measurable voltage. Hence the SiPM's output is represented by an analog pulse. Since SiPMs do not have a built-in photon counter, two multi-photon readout methods were developed to count SiPM pulses: voltage thresholding and summation output \cite{Long1}. The voltage thresholding method is achieved by counting pulses above the voltage threshold. In contrast, the summation output method is achieved by sums of the pulses and has a higher dynamic range than the voltage thresholding method \cite{Long1}.

Recently, to achieve higher data rates, the summation output method has been evaluated with a larger array of SiPMs. For example, a commercially available $3.07\times~3.07$~$mm^2$  SiPM has been used to achieve OOK data rates up to 2 Gbps at BER of $10^{-3}$ with a sensitivity of -29~dBm by using decision feedback equalisation (DFE) \cite{Zubair1}. Another experimental result shows the data rate was improved to 5~Gbps with the orthogonal frequency division multiplexing (OFDM) method at 9~$\mu W$, which is equivalent to -20.5~dBm \cite{Shenjie}. Apart from the VLC application, the scenario of underwater wireless optical communication (UWOC) has also been investigated. Recent work in this field showed that by applying the summation readout method with DFE, the data rate could reach 1~Gbps \cite{Long2}, which is higher than the 7.9~Mbps by applying the photon counting. An additional study showed the potential of using SiPMs through the implementation of a 6x3 multiple-input and multiple-output (MIMO) scheme, which utilized single photon counting to achieve a data rate of 1~Mbps over an underwater link\cite{Jinjia}. More recent research demonstrates the potential of using CMOS low-power consumption circuits \cite{Goll} and satellite-based optical communication \cite{Alexander}. 

The previous literature showed that the summation method with DFE has the advantage of a higher OOK data rate \cite{Zubair1, Shenjie, Long2}. However, at data rates when the pulses are countable and the output pulses do not overlap, the thresholding method for single photon counting is still necessary, especially in scenarios where the incident optical power of the receiver is extremely low. In addition, the data rate cannot be reduced flexibly, hence operating at a specific data rate that is expected to be twice the recovery time in OOK \cite{Danial}.

\section{SiPM Characterisation}

\subsection{SiPM Background}

In this work, the detector is a commercially available $1~mm^2$ C-Series SiPM from On Semiconductor \cite{Onsemi}. The technical parameters for the SiPM are listed in Table \ref{tab: SiPM Parameters}. 


\begin{table}[t]
  \centering
  \renewcommand{\arraystretch}{1.2}
  \caption{Summary of key parameters of the SiPM}
  \label{tab: SiPM Parameters}
   \begin{tabular}{@{}p{5cm}ll@{}}
    \toprule
    \toprule
    Parameters                 & Value    \\ \Cline{1-2}{1pt}
    Area                       & 1 \(\text{mm}^2\)  \\ 
    Microcell size             & 10 $\mu$m \\ 
    Number of SPADs            & 2880   \\ 
    Typical break down voltage & 24.5 V \\ 
    Typical dark count rate    & 30 kHz \\ 
    Fill factor                & 28\%   \\ 
    Peak wavelength            & 420 nm \\ 
    & 14\% at 420 nm\\
    \multirow{-2}{*}{PDE at 2.5 V overvoltage} & 3.6\% at 620 nm \\
    \toprule
    \toprule
  \end{tabular}
\end{table}

When a photon is absorbed by the junction, the photo-current generated by the avalanche event flows through a load resistor to convert the total avalanche photo-current into a measurable voltage. After the amplification, this voltage is an analog pulse, which represents the received photon count. The incident photon energy is characterized by the wavelength of the light source and is calculated using Planck's equation\cite{Griffiths}. Due to the quantum efficiency of silicon, only a certain percentage of the incident photons are detected and converted into analog pulses. This is usually described with photon detection efficiency (PDE) and is defined as \cite{Stewart}:

\begin{equation} 
PDE\left(  \lambda, V\right) =\eta\left( \lambda \right)\times\varepsilon\left( V \right)\times F
\end{equation}

Where $\eta\left( \lambda \right)$ is the quantum efficiency of the p-n junction at a given wavelength, $F$ is the fill factor of the SiPM, and $\varepsilon\left( V \right)$ is the avalanche initiation probability, which is a function of the applied bias V. As the PDE already accounts for the fill factor detailed in Table \ref{tab: SiPM Parameters}, it is essential to calibrate the PDE at the specified wavelength with the datasheet to guarantee that the detector receives the correct number of detected photons. Then, the detected photon count is written as:

\begin{equation}
\label{eq: lambda_s}
\lambda_{s} = \frac{E_{total}}{E_{photon}}\times PDE\left(  \lambda, V\right)
\end{equation}

Where $E_{total}$ is the total energy of the SiPM receiving optical power during a bit time, and $E_{photon}$ is the energy for a single photon. After considering the PDE, the detected photons are related to the incident photons, reflecting the power or irradiance needed by the SiPM. Since SiPM detects single photons, its photon statistics are based on Poisson (shot noise) statistics. The limit defined by photon statistics is known as the quantum limit, and it is also referred to as the Poisson limit since the Poisson statistic is typically employed in photon counting scenarios. 

When a random bit sequence is transmitted, the probability of error (PE) is estimated using the following equation \cite{Gagliardi}:

\begin{equation}
\label{eq: PE}
\begin{split}
    PE=\frac{1}{2} \times \sum_{k=0}^{n_{T}} \frac{\left( \lambda_{s} + \lambda_{b}\right)^{k}}{k!}e^{-\left( \lambda_{s} +\lambda_{b} \right)} \\+ \frac{1}{2} \times \sum_{k= n_{T}}^{\infty} \frac{\left( \lambda_{b} \right) ^ {k}}{k!}e^{-\left( \lambda_{b} \right)} 
\end{split}   
\end{equation}

Where $\lambda_{s}$ is the number of detected signal photons per bit time and $\lambda_{b}$ is the number of detected background photons per bit time, which is calculated from (\ref{eq: lambda_s}). $k$ is the counter of the photon event. The detected background photons are counted when 0s are transmitted. However, when 1s are transmitted, the detected photon count includes both signal and background photons. The detected background photons include dark counts of the SiPM and light leakage in the measurement setup. The $n_{T}$ is an integer decision threshold to determine 0s and 1s, and it is always selected to achieve the best PE. Based on (\ref{eq: PE}), a PE of $10^{-3}$ can be achieved by an average of 6.2 detected photons if there is no background photon. When the background photons are presented, more $\lambda_{s}$ is required to maintain the PE of $10^{-3}$, hence, a power penalty is considered. Once the practical BER of $10^{-3}$ is achieved, FEC could be used to further improve the BER \cite{Tychopoulos}.

\subsection{Experimental Configuration}
To demonstrate the real-time optical receiver with SiPM, an AMD/Xilinx PYNQ-Z1 evaluation board with Zynq-7000 SoC XC7Z020-1CLG400C FPGA was chosen as the platform to characterize SiPM output pulses. Considering the peripheral connection between SiPM output and FPGA board, the peripheral module (PMOD) interface provided by the board was used \cite{Singh}. The PMOD interface was developed by Digilent Inc. for the low frequency, low I/O peripheral connections. The expected bandwidth of the PMOD interface is tens of megahertz. Since the digital signal characteristics are not specified, the maximum speed for digital SiPM pulse detection was evaluated. 

\begin{figure} 
    \centering
\subfloat[\label{fig: PMOD setup}]{
    \includegraphics[width=0.48\linewidth]{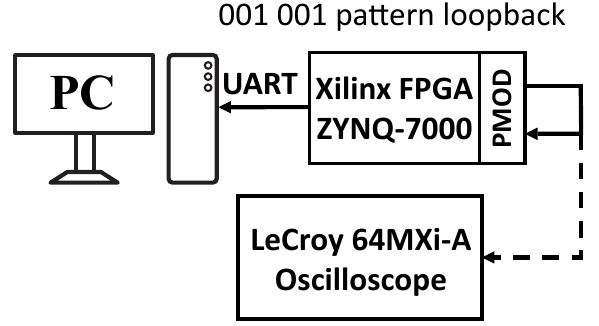}}
    \hfill
\subfloat[\label{fig: PMOD pulse width}]{
    \includegraphics[width=0.48\linewidth]{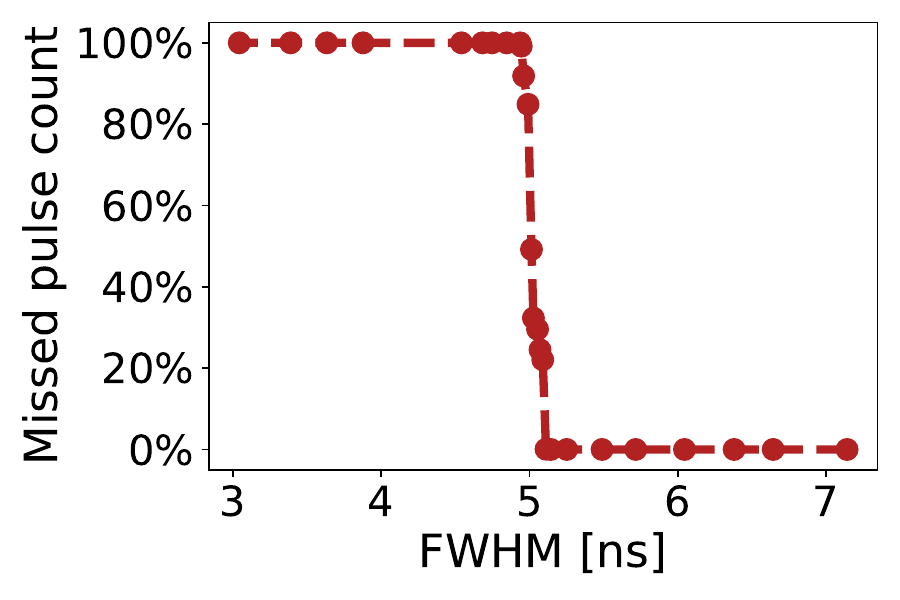}}
 
  \caption{Measuring the minimum digital pulse width which could be detected by the PMOD connector using a loop-back test. (a) experimental setup (b) the minimum pulse width of the detected pulse without error.}
  \label{fig: PMOD} 
\end{figure}

Fig.~\ref{fig: PMOD}\subref{fig: PMOD setup} shows the experimental setup for measuring the minimum digital pulse width, which could be detected by the PMOD connector using a loop-back test. Prior to the test, a fixed 001001 pattern signal was generated using the FPGA's internal phase lock loop (PLL). This signal was then connected through the PMOD connector and calibrated on a 600~MHz 10~GS/s LeCroy 64MXi-A oscilloscope. During the loop-back test, the wire connection between PMOD connectors was as short as possible to ensure the best signal integrity performance. The PC received the pulse counts in every 1 second through the universal asynchronous receiver-transmitter (UART) in real-time. As the clock frequency of the generated pattern increased, the average full width half maximum (FWHM) of each pulse within the 001001 patterns was measured during a 10~ms waveform capture. 

Fig.~\ref{fig: PMOD}\subref{fig: PMOD pulse width} shows the missed pulse counts during the pulse width changes. It could be observed that the minimum pulse width of the PMOD connector detected without error is approximately 5~ns. In this condition, the pulse amplitude was tested at approximately 3.3~V, which meets the low voltage complementary metal-oxide-semiconductor (LVCMOS) 3.3~V and low voltage transistor-transistor logic (LVTTL) 3.3~V standards. Therefore, the expected SiPM pulse width to demonstrate the real-time receiver should be longer than 5~ns.

\begin{figure}
  \begin{center}
  \includegraphics[width=3.2in]{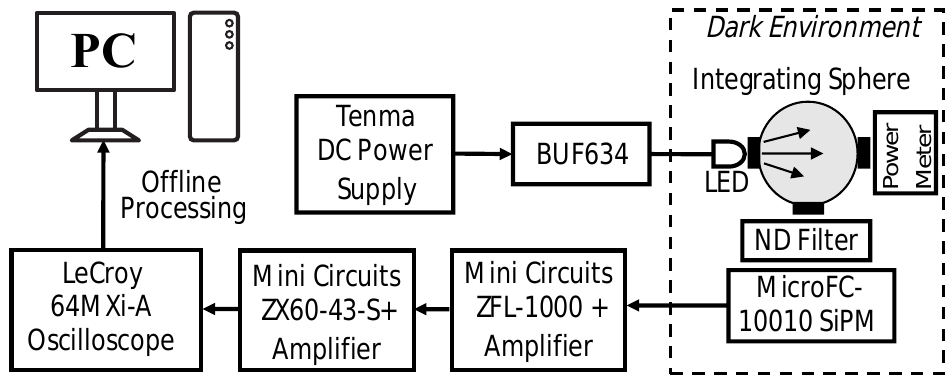}\\
  \caption{Experimental setup for characterization of standard and fast outputs of the SiPM.}
  \label{fig: offline setup}
  \end{center}
\end{figure}

The offline experimental setup shown in Fig.~\ref{fig: offline setup} was then built to characterize the SiPM output before the implementation of the real-time setup. The transmitter consists of a Tenma DC power supply, which powers a TI voltage buffer BUF634 to drive a 626~nm light-emitting diode (LED). The optical intensity is controlled by the bias voltage of the direct current (DC) power supply. 

On the receiver side, since the receiving light intensity of SiPM needed to be accurately measured, an integrating sphere Thorlabs IS200-4 was used to generate equal photon flux among the LED, MicroFC-10010 SiPM, and optical power meter. As the LED light enters the integrating sphere, it experiences numerous diffusions and reflections, ultimately achieving an even distribution over the entire inner surface of the sphere. To reduce the light intensity reaching the SiPM, a Thorlabs neutral density (ND) filter with an optical density of 20 is used. This attenuation in intensity allows the SiPM to work within its linear response range and enhances the measurement precision. The SiPM output signal was then amplified by two Mini-Circuit amplifier blocks: ZX60-43-s+ and ZFL-1000+. The average gain and maximum bandwidth of ZX60-43-s+ is 18.6~dB and 4~GHz, and ZFL-1000+ is 17~dB and 1~GHz. After the amplification, the SiPM output pulse was captured by the LeCroy oscilloscope. The internal memory of the oscilloscope allowed a maximum of 10~ms waveform capture each time. When the PC remotely captures and saves sufficient waveform data from the oscilloscope, the PC analyses the data samples.

\subsection{SiPM Pluses Characterization}

The traditional SiPM readout depends on the resistor between the anode and cathode, referred to as the standard output. In addition to the resistor, a capacitor coupling the output of each microcell was developed to obtain a shorter pulse, referred to as the fast output. The standard output is commonly available in all commercial SiPMs, rather the fast output has been identified only for on-semi SiPMs after comparing the fast and standard outputs of SiPM 10010. In order to make the post-readout circuit applicable to the majority of commercially available SiPMs, the paper primarily aimed to evaluate the standard output. Based on the setup in Fig.~\ref{fig: offline setup}, the pulse capture and statistic were performed to evaluate whether both outputs were compatible with PMOD connectors.

\begin{figure} 
    \centering
\subfloat[\label{fig: fast pulse}]{
    \includegraphics[width=0.48\linewidth]{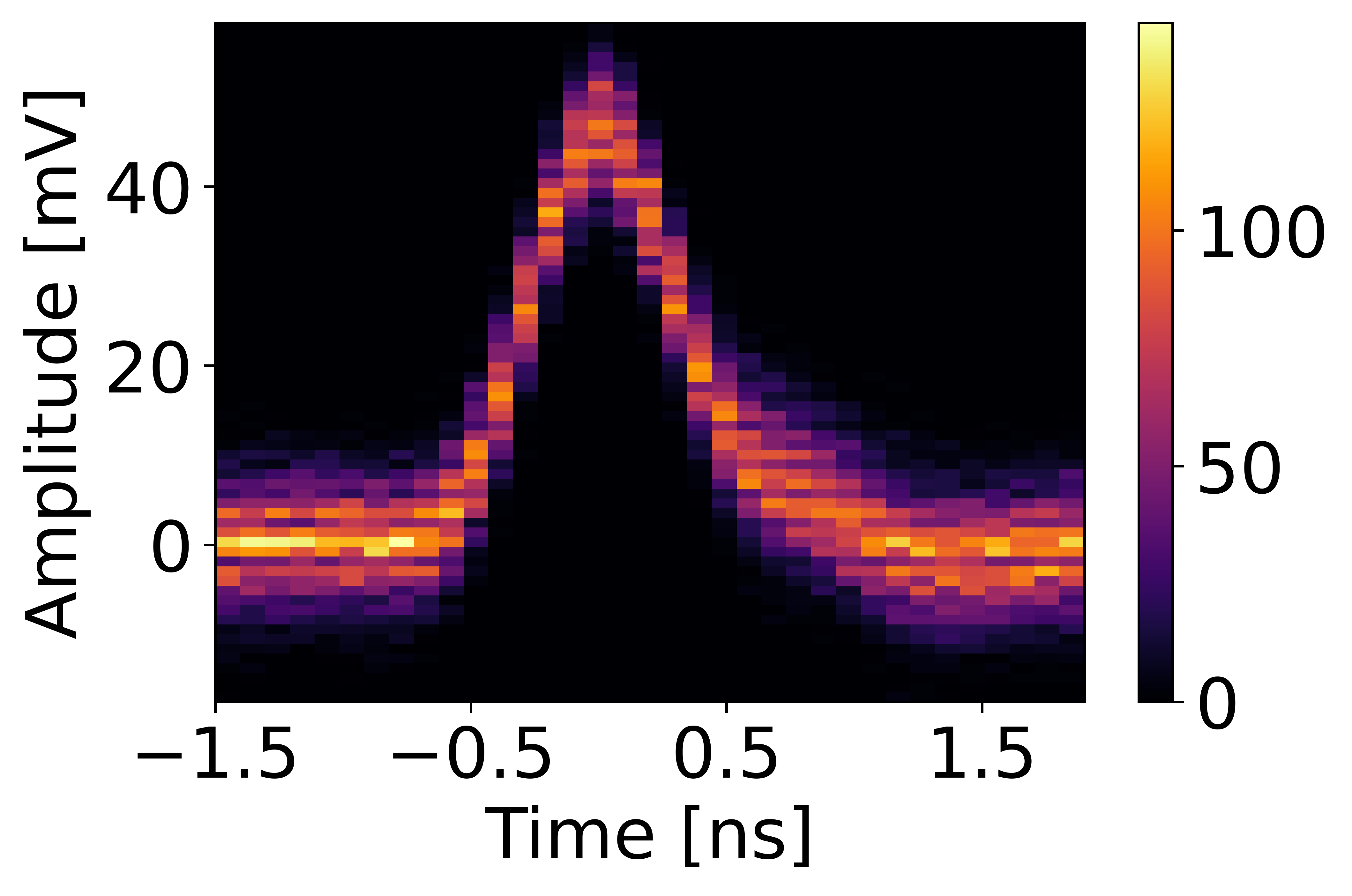}}
    \hfill
\subfloat[\label{fig: standard pulse}]{
    \includegraphics[width=0.48\linewidth]{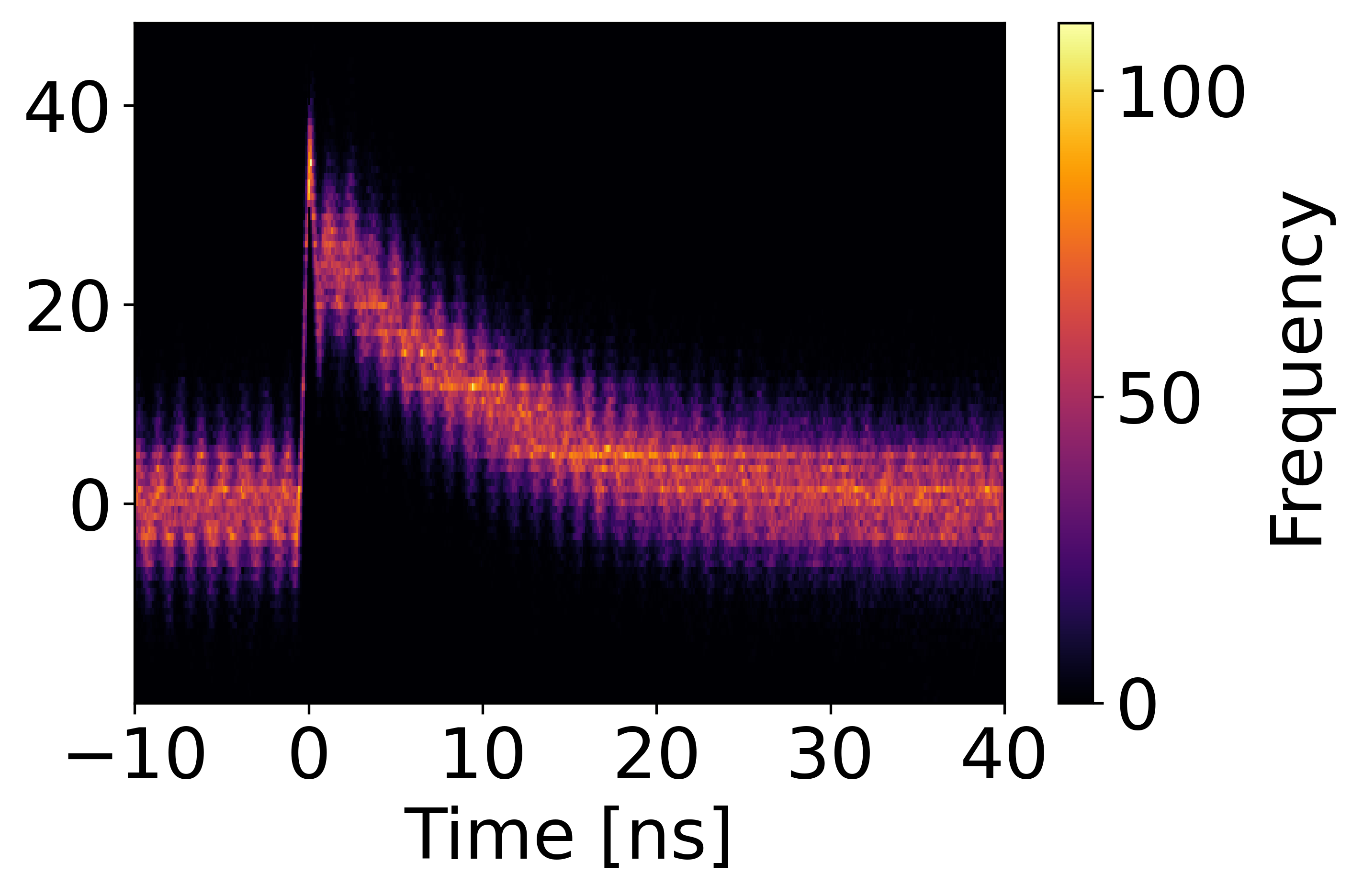}}
 
  \caption{Histogram of 1000 captured output voltage pulses of the SiPM  (a) fast output with average FWHM of approximately 1~ns. (b) standard output pulses with an average FWHM of approximately 8~ns.}
  \label{fig: fast and standard pulse} 
\end{figure}

Fig.~\ref{fig: fast and standard pulse} shows the offline processed pulse shape of a single SiPM output pulse for fast and standard outputs, with 1000 accumulated pulses displayed in the colored histogram. In Fig.~\ref{fig: fast and standard pulse}\subref{fig: fast pulse}, it was observed that the amplitude of the fast output pulse was around 50~mV when the SiPM was biased at 27.5~V. The amplitude of the standard output pulse was approximately 40~mV, which was lower than the fast output pulse, shown in Fig.~\ref{fig: fast and standard pulse}\subref{fig: standard pulse}. The FWHM of the fast output pulse was shorter than the standard output pulse, 1~ns versus 8~ns, which means the bandwidth of the PMOD connector is insufficient to read out the fast output pulses. In this case, the standard output was chosen to perform the real-time connection to FPGA hardware. Moreover, the standard output is more commonly available in commercial SiPMs after comparing the fast and standard outputs of SiPM 10010.

To convert the analog pulse of the standard output to a readable digital pulse, a robust thresholding method was developed based on the behavior of the electronic op-amp comparator with hysteresis. In a comparator circuit, hysteresis is applied by using positive feedback to add a control voltage between the upper and lower threshold voltages. This hysteresis minimizes the electronic noise and improves decision accuracy. Although the offline hysteresis was done to determine the dynamic range and maximum photon count rate before the real-time system, the hysteresis of the real-time system needs to determine again due to the new configuration, which has a different noise floor and pulse amplitude. Hence, the hysteresis was applied in both offline and real-time processing.
\begin{figure}[t]
    \centering
    \includegraphics[width=3.2in]{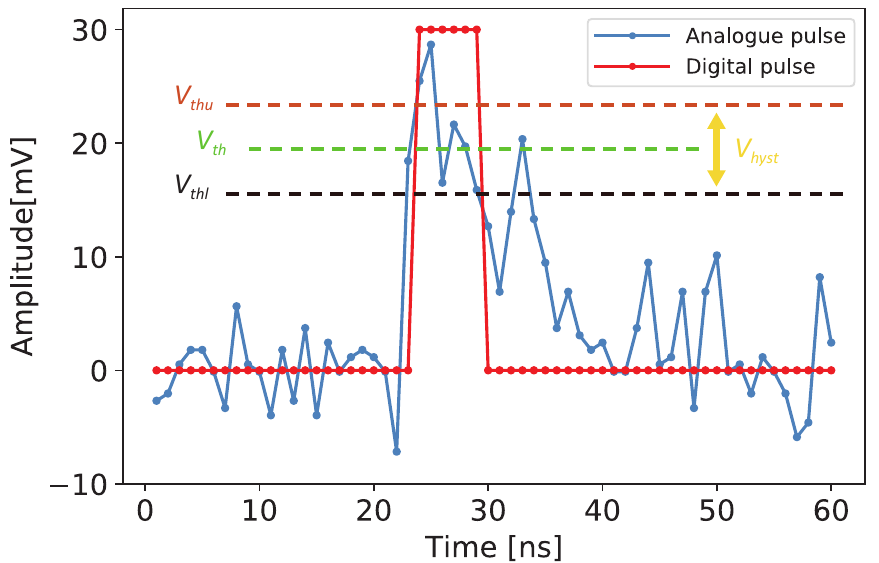}
    \caption{The principle of SiPM's analog to digital pulse conversion by using the hysteresis method. $V_{th}$ is the threshold voltage.}
    \label{fig: hysteresis}
\end{figure}

Fig.~\ref{fig: hysteresis} illustrates the offline processed analog-to-digital conversion (ADC) with a single pulse waveform captured by the oscilloscope with 1~GS/s sampling rate. From this figure, the pulse's falling edge has a long tail with a relatively large noise compared with the pulse amplitude. Under this circumstance, the wrong decision would happen if threshold voltage $V_{th}$  was lower than the noise amplitude. To develop a robust thresholding method, a 5~mV hysteresis voltage $V_{hyst}$ was added based on the amplitude of DC noise shown in Fig.~\ref{fig: fast and standard pulse}\subref{fig: standard pulse}. When the pulse sample amplitude is detected higher than $V_{thu}$, the next series of samples will be converted to a digital 1s until the subsequent samples are detected lower than $V_{thl}$.

\subsection{Dynamic Range of Offline Setup}

\begin{figure}[t]
    \centering
\subfloat[\label{fig: offline threshold sweeping}]{
    \includegraphics[width=0.92\linewidth]{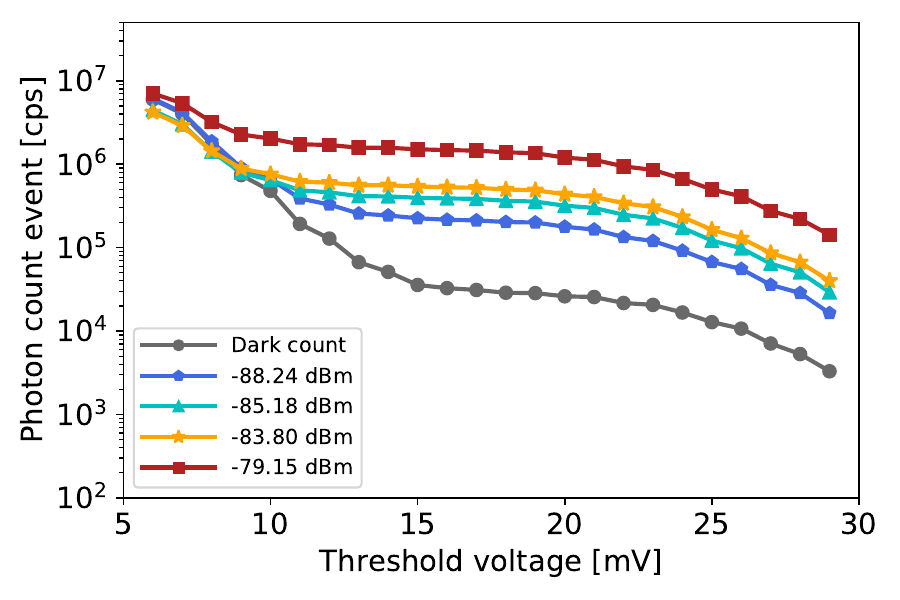}}
    \hfill
\subfloat[\label{fig: dynamic range}]{
    \includegraphics[width=0.92\linewidth]{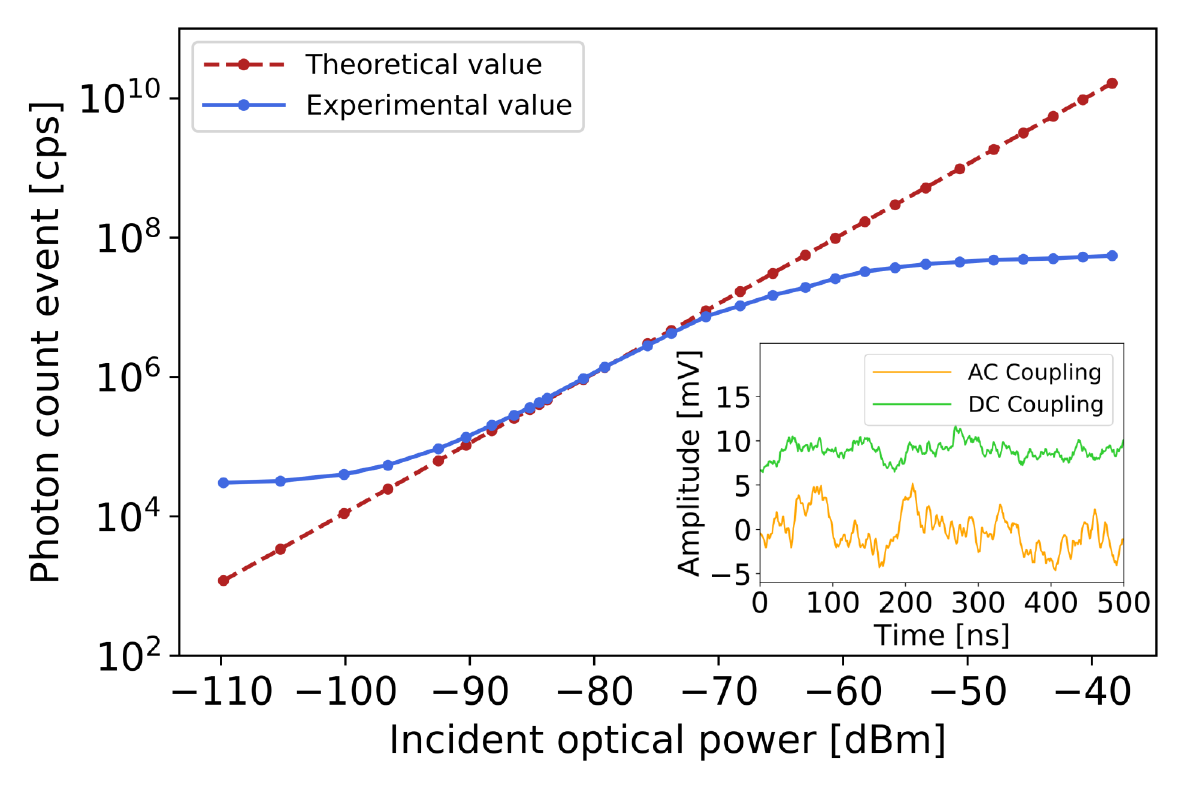}}
    \hfill
  \caption{ The noise floor and dynamic range measurement of offline setup. (a) The output detected photon count when the threshold voltage $V_{th}$ varies from 6~mV to 29~mV. The detected dark count is approximately 30~kcps when the $V_{th}$ was configured as 18~mV. (b) The detected photon count event as a function of incident optical power compared with the expected theoretical linear response using the selected $V_{th}=18~mV$. The inset plot shows the difference between DC and AC coupled SiPM output under -43.4~dBm incident optical power when the SiPM was saturated.}
  \label{fig: offline result} 
\end{figure}

To find the suitable threshold voltage $V_{th}$, a threshold sweeping was performed with a step of 1~mV. Fig.~\ref{fig: offline result}\subref{fig: offline threshold sweeping} shows the counting result of the sweeping threshold voltage under different SiPM received optical power. As expected, the detected photon count per second increased with the increasing optical power. Additionally, at higher thresholds, independent counting was observed when the $V_{hyst}$ range was above the electrical noise floor but below the pulse amplitude. If the $V_{hyst}$ is increased to be higher than the maximum pulse amplitude, the pulse count will be lost. Hence, a $V_{th}$ of 18~mV was chosen for all optical power measurements.

The detected photon count event per second at different optical power after employing threshold voltage $V_{th}$ was shown in Fig.~\ref{fig: offline result}\subref{fig: dynamic range}. In this figure, the linear theoretical response was calculated from (\ref{eq: lambda_s}) when the calibrated effective PDE of 3.6\% was applied. It was observed that when the optical power was lower, the dark count played a primary role in the detected photon count, which was at 30~kcps. When the incident power increases, the detected photon count response is close to linear hence approaching the Poisson limit. As the incident optical power is further increased, the detected photon count deviates from the theoretical incident photon count. This is because the bit time required for a 1~Mbps data rate is much longer than the pulse width of SiPM. As a result, the observed non-linearity is caused by the effects of overlapping output pulses instead of the conventional cause of intersymbol interference (ISI) \cite{Salvatore}.

In the presence of non-linear output, the results show a relatively constant detected photon count, which is due to the coupling of the SiPM output pulse to the analog signal chain. The inset in Fig.~\ref{fig: offline result}\subref{fig: dynamic range} shows the waveform difference between alternating coupling (AC) and direct coupling (DC) SiPM output captured from the oscilloscope. Regarding DC, the voltage thresholding method was not sustained due to the increasing offset of the pulse waveform when the detected photon counts are higher, resulting in a slight increase in detected photon count. In contrast, the output of AC oscillates around 0~V, hence a constant detected photon count.  Since the amplifiers in the signal chain are AC, a constant detected photon count event is expected.


\section{Experimental BER Measurement}

\subsection{Setup}

\begin{figure}[t]
    \centering
    \includegraphics[width=3.4in]{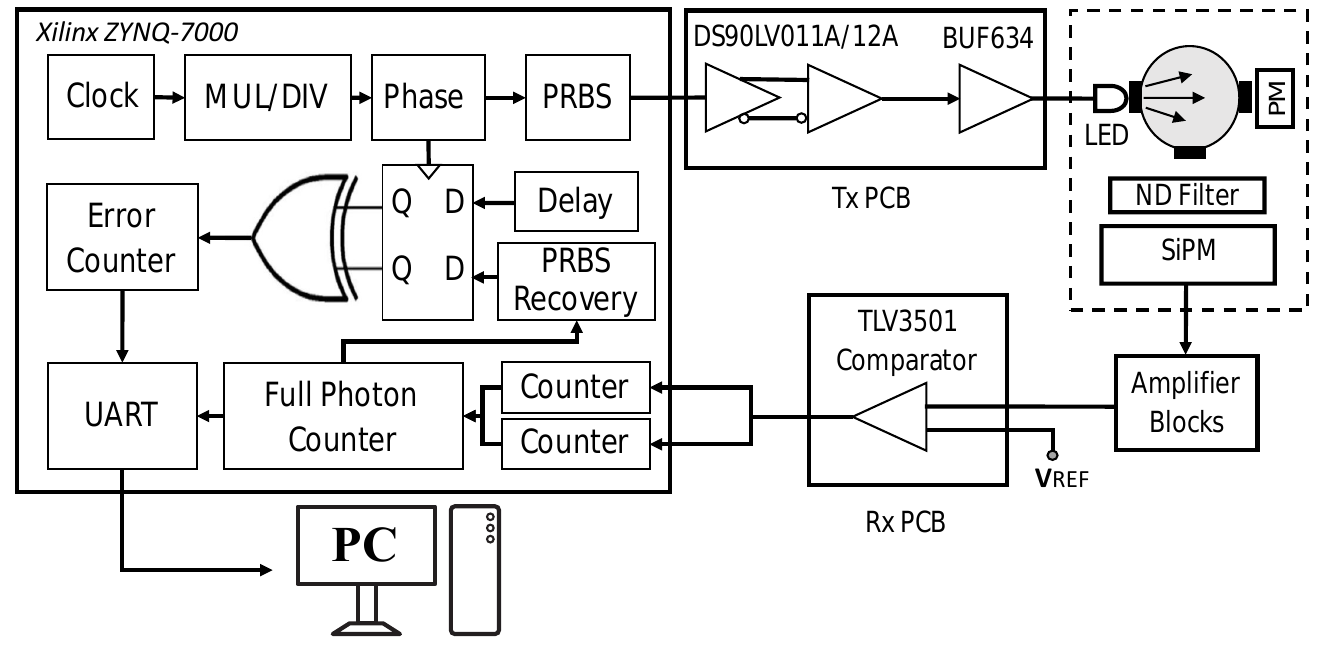}
    \caption{The real-time experimental setup for SiPM's BER test.}
    \label{fig: realtime setup}
\end{figure}

Fig.~\ref{fig: realtime setup} shows the recently developed experimental setup utilized to investigate the real-time detected photon count for varying incident light intensities and their corresponding BER \cite{Yangchun}. The digital parts of the system were developed based on the FPGA board, including clock generation, a pseudo-random bit stream (PRBS) generator based on linear feedback shift register (LFSR), an exclusive-OR (XOR) gate comparison block, counters to count errors and detected photons, and a UART interface sending the counter data to a PC. To achieve reasonable signal integrity, a differential transmission was employed to connect the LED to the FPGA board.

The transmitter (Tx) printed circuit board (PCB) included a DS90LV011A low-voltage differential signaling (LVDS) differential line driver, a DS90LV012A LVDS differential line receiver, and a voltage buffer to drive the LED. The receiver (Rx) part, considering the low amplitude of the SiPM pulse, used three amplifier blocks, including one Mini-Circuits ZFL-1000+ and two Mini-Circuits ZX60-43-S+, which were connected in series to amplify the SiPM pulse amplitude to approximately 400~mV. After the amplification, a TLV3501 comparator with a 4.5~ns response time designed on the Rx PCB was used to convert the analog SiPM pulse to a readable digital pulse by FPGA. The threshold voltage $V_{th}$ for the comparator in the real-time setup needs to be selected again when there are modifications to the system since it is determined by the level of noise. In the real-time system, The increase in noise floor caused by the introduction of an additional ZX60-43-S+ component means that the previously chosen value of 18~mV for $V_{th}$ in offline processing is unsuitable and must be updated.

\begin{figure}[t]
    \centering
\subfloat[\label{fig: comparator threshold sweeping}]{
    \includegraphics[width=0.92\linewidth]{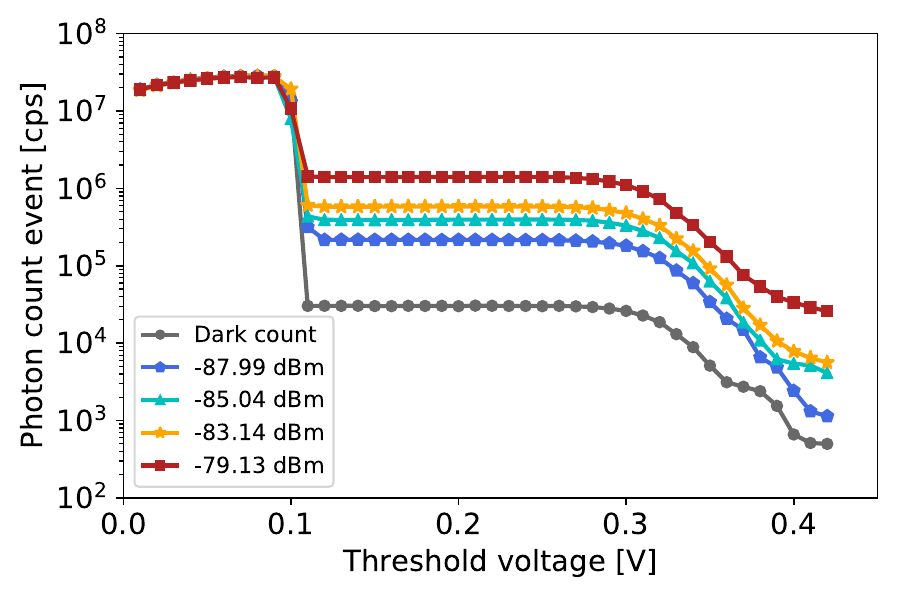}}
    \hfill
\subfloat[\label{fig: DR_offline_realtime}]{
    \includegraphics[width=0.92\linewidth]{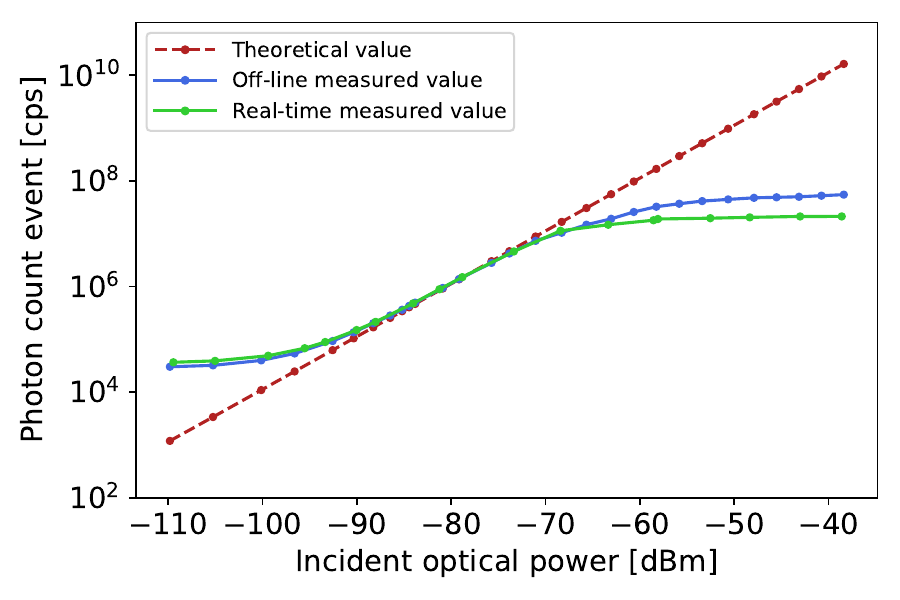}}
    \hfill
  \caption{ The noise floor and dynamic range measurement of real-time setup. (a) The output detected photon count when the comparator's threshold voltage varies from 0~V to 0.42~V (b) The comparison of dynamic range between offline setup and real-time setup.}
  \label{fig: realtime result} 
\end{figure}

To maintain a low power consumption, an asynchronous SiPM pulses counter was implemented in the FPGA. This asynchronous counter was triggered by the rising edge of the random pulses from the SiPM so that it did not require a high-frequency clock for sampling the input pulses. However, the decision of bit 0s and bit 1s for each time slot is still required to be made by selected digital threshold $n_{T}$. To achieve uninterrupted real-time counting, two inter-leaved counters were implemented. When the digital pulses during one bit time arrived at the FPGA, only one counter received and counted the pulse at any time. Then, the other counter would save the counting result to the full photon counter for the accumulation of whole bit counting information to calculate the BER. When the counting information was saved, the counter would be reset and wait for pulses for the next bit. Therefore, continuous pulse counting for each bit without any dead time was achieved.

\subsection{Dynamic Range of Real-time Setup}

Before the BER measurement, the analog threshold voltage $V_{th}$ of the comparator was swept to find the region where counts were independent of the $V_{th}$. From Fig.~\ref{fig: realtime result}\subref{fig: comparator threshold sweeping}, the real-time dark counts remain 30~kcps when the threshold is swept from 0.1~V to 0.3~V. These output counts matched the typical dark counts from the SiPM's datasheet. When the SiPM incident optical power was high, the counting results also increased and maintained the real detecting photons. Based on this observation, the analog threshold voltage was configured as 0.2~V. This is higher than offline processing $V_{th}$ due to the addition of the third amplifier.

Fig.~\ref{fig: realtime result}\subref{fig: DR_offline_realtime} presents a comparison of the dynamic range measurement between the offline setup and the real-time setup. The dark count was measured at 28~kcps and 35~kcps for the offline and real-time setups, respectively, which is consistent with the typical value of 30~kcps specified in the datasheet. Although the linear region of the two measurements matches well at higher incident optical power, the detected photon count event is lower for the real-time setup than the offline setup when the SiPM becomes saturated. This is due to the bandwidth limitation of the comparator and PMOD connector, which were not included in the offline setup compared to the oscilloscope channel bandwidth of 600 MHz. The measurement of SiPM output on an oscilloscope in the offline setup enables debugging and characterization, and while this is an intermediate step, it represents a transitional phase toward the final objective of real-time implementation.

\subsection{BER Result}
\begin{figure}[t]
    \centering
\subfloat[\label{fig: FPGA threshold sweeping}]{
    \includegraphics[width=0.92\linewidth]{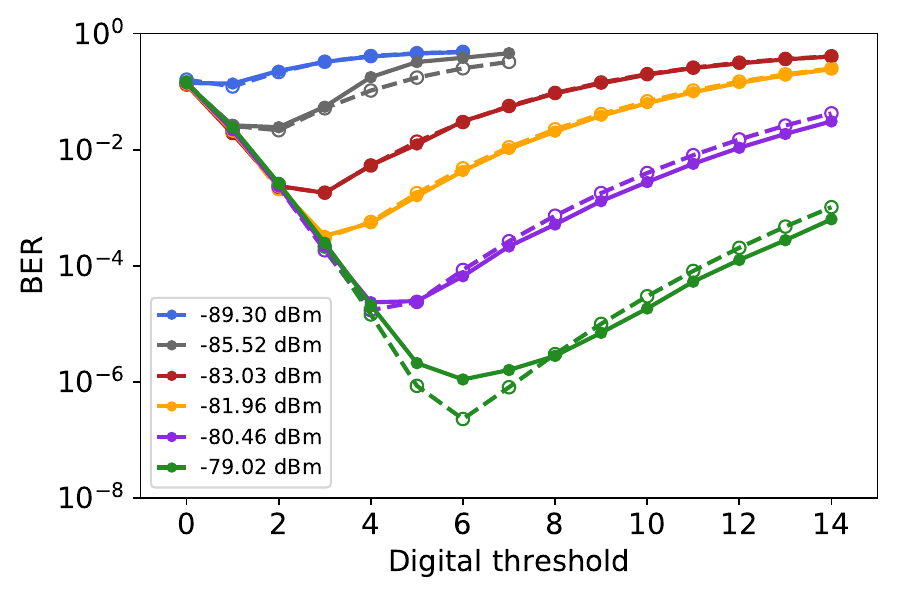}}
    \hfill
\subfloat[\label{fig: BER vs. optical power}]{
    \includegraphics[width=0.92\linewidth]{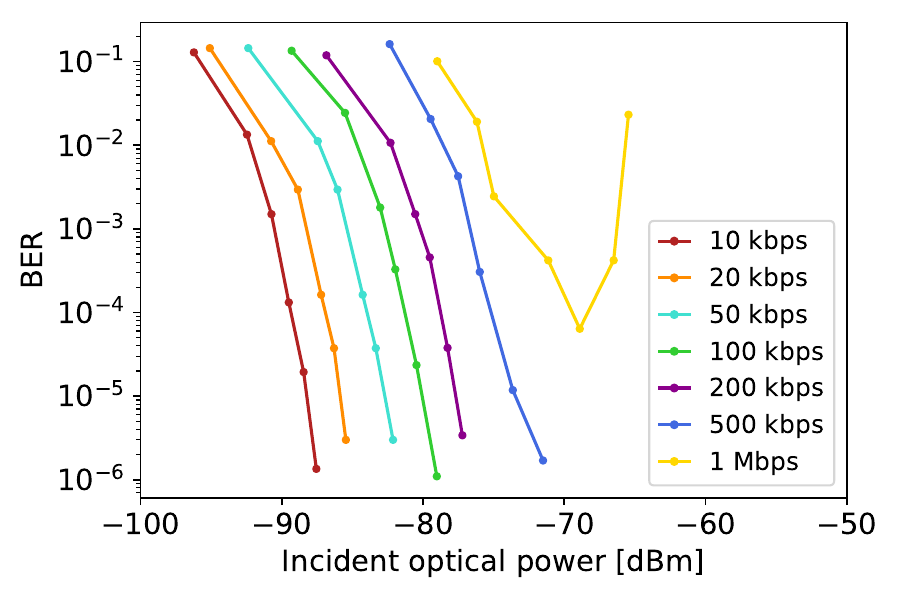}}
    \hfill
  \caption{The results for the experimental real-time setup (a) The measured BER result when the digital threshold $n_{T}$ varies from 0 to 15 at 100~kbps. The dashed line shows the Possion limit result from (3). (b) Experimental BER versus incident optical power from the data rate 10~kbps to 1~Mbps. The collection time for achieving a BER below $10^{-3}$ is 1 second. For achieving a BER of $10^{-4}$, the collection time is 10 seconds. For achieving BERs of $10^{-5}$ and $10^{-6}$, the collection times are 100 seconds.}
  \label{fig: BER result} 
\end{figure}

\begin{table*}[t]
\centering
\renewcommand{\arraystretch}{1.2}
\caption{The power consumption of receiver component}
\label{tab: Experimental result}

\begin{tabular}{cccccc}
\toprule
\toprule
\begin{tabular}[c]{@{}c@{}}Name of the \\ receiver component\end{tabular} & Quantity             & Supply voltage          & Test condition                                                                            & Measured current & Power consumption \\ \Cline{1-6}{1pt}
\multirow{5}{*}{SiPM 10010}                                               & \multirow{5}{*}{1} & \multirow{5}{*}{27.5 V} & \begin{tabular}[c]{@{}c@{}}Optical power 0.85 pW\\  @10 kbps, $10^{-3}$ BER\end{tabular}  & 8.06 nA          & 0.22 $\mu$W           \\ \Cline{4-6}{0.01pt}
                                                                          &                    &                         & \begin{tabular}[c]{@{}c@{}}Optical power 4.98 pW \\ @100 kbps, $10^{-3}$ BER\end{tabular} & 46.56 nA         & 1.28 $\mu$W           \\ \Cline{4-6}{0.01pt} 
                                                                          &                    &                         & \begin{tabular}[c]{@{}c@{}}Optical power 31.78 pW \\ @1 Mbps, $10^{-3}$ BER\end{tabular}   & 291.08 nA        & 8.06 $\mu$W           \\ \Cline{1-6}{1pt}
Amplifier ZX60-43+                                                        & 2                  & 5 V                     & \multirow{3}{*}{SiPM under incident light}                                                   & 87 mA            & 0.87 W   \\ \Cline{1-3}{1pt} \Cline{5-6}{1pt} 
Amplifier ZFL1000+                                                        & 1                  & 15 V                    &                                                                                          & 74 mA            & 1.11 W   \\ \Cline{1-3}{1pt} \Cline{5-6}{1pt} 
Comparator TLV 3501                                                       & 1                  & $ \pm $ 3.3 V                   &                                                                                          & 3 mA             & 19.8 mW           \\ \Cline{1-6}{1pt}
\multirow{3}{*}{FPGA Zynq 7000}                                           & \multirow{3}{*}{1} & \multirow{3}{*}{12 V}   & \begin{tabular}[c]{@{}c@{}}Impelmentation of \\ transmitter\end{tabular}                 & 154 mA           & 1.848 W           \\ \Cline{4-6}{0.01pt} 
                                                                          &                    &                         & \begin{tabular}[c]{@{}c@{}}Impelmentation of \\ transmitter and receiver\end{tabular}    & 157 mA           & 1.884 W           \\ 
\toprule
\toprule
\end{tabular}
\end{table*}

\begin{figure}[t]
    \centering
    \includegraphics[width=3.2in]{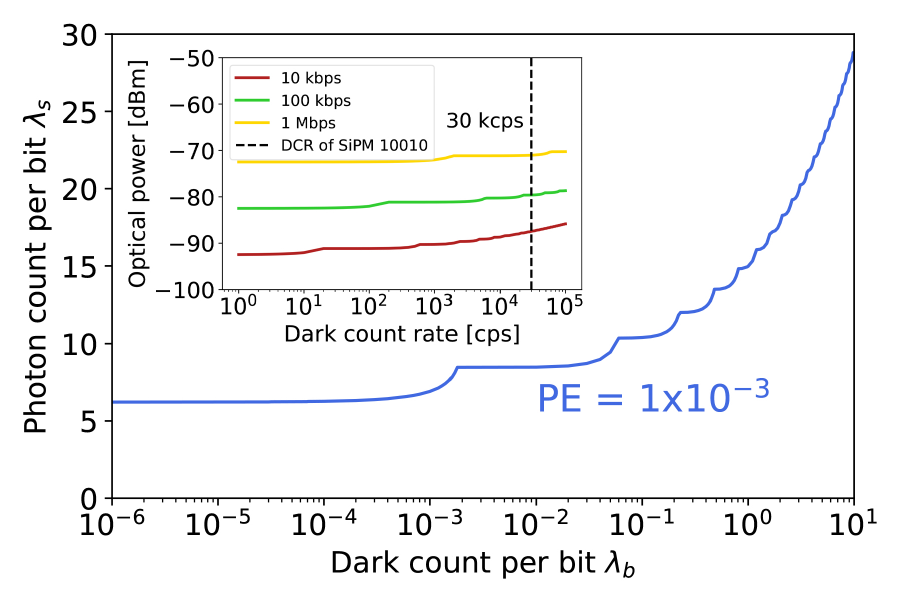}
    \caption{The theoretical detected photon count versus dark count per bit under a PE of $10^{-3}$ computed by (3). The inset plot shows the required theoretical incident optical power to achieve a PE of $10^{-3}$, considering various dark count rates when data transfer rates are under 1 Mbps. The theoretical incident optical power was calculated by the theoretical detected photon count and effective PDE.}
    \label{fig: data rate vs. counts per bit}
\end{figure}

\begin{figure*}[t]
\centering
\subfloat[\label{fig: G1_BW500}]{
    \includegraphics[width=0.34\linewidth]{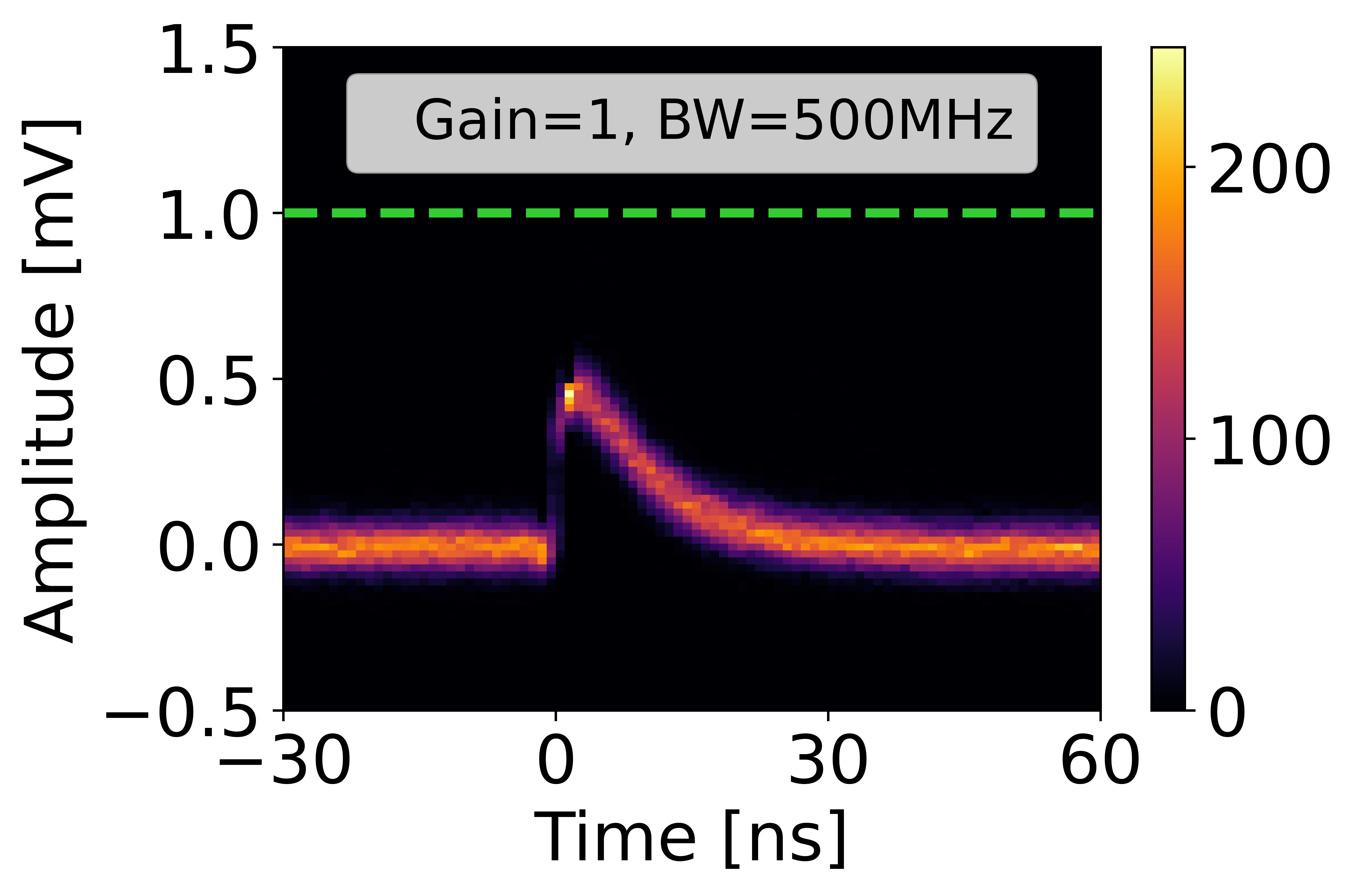}}
    \hfill 
\subfloat[\label{fig: G10_BW50}]{
    \includegraphics[width=0.30\linewidth]{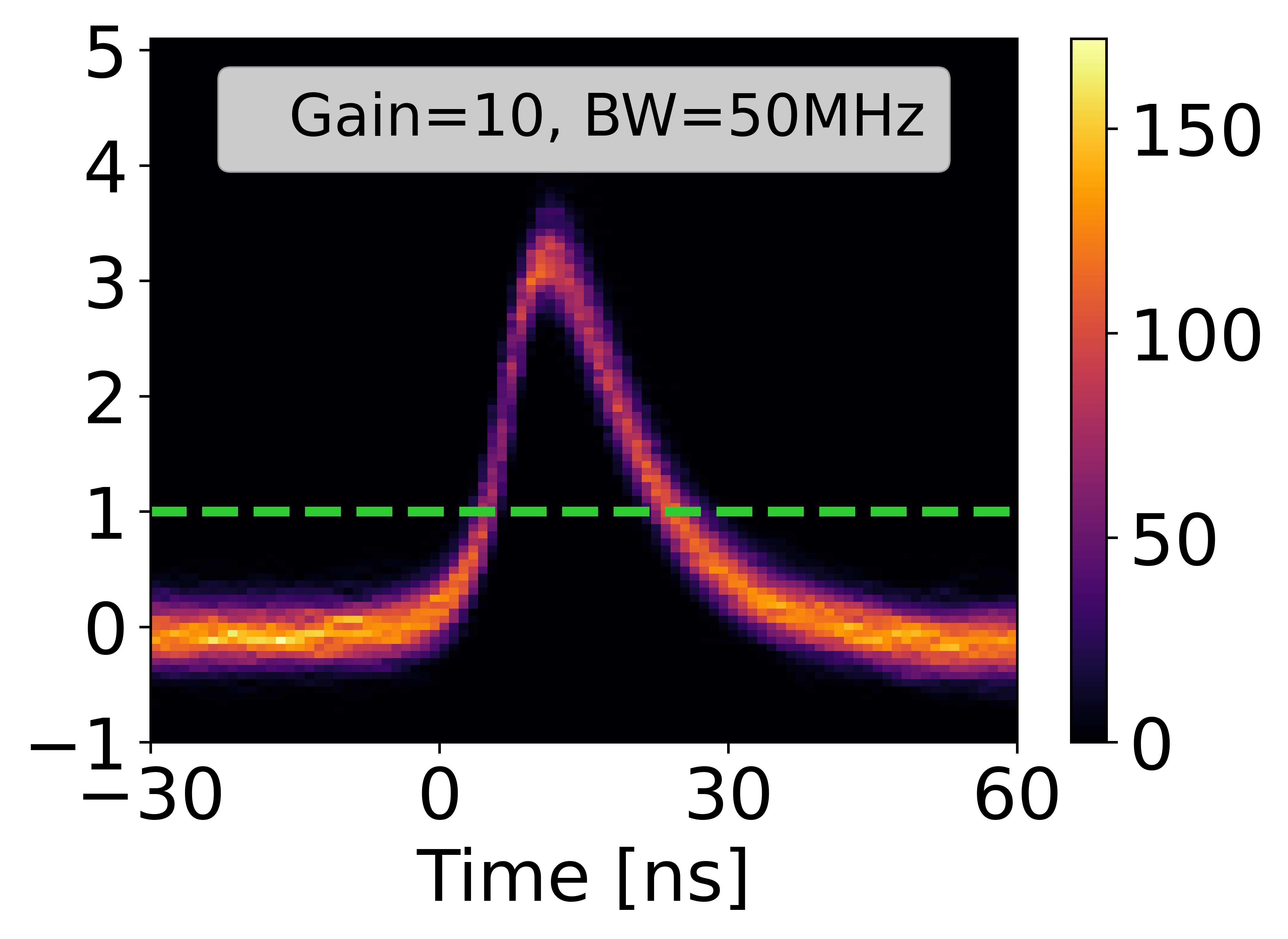}}  
    \hfill 
\subfloat[\label{fig: G500_BW1}]{
    \includegraphics[width=0.31\linewidth]{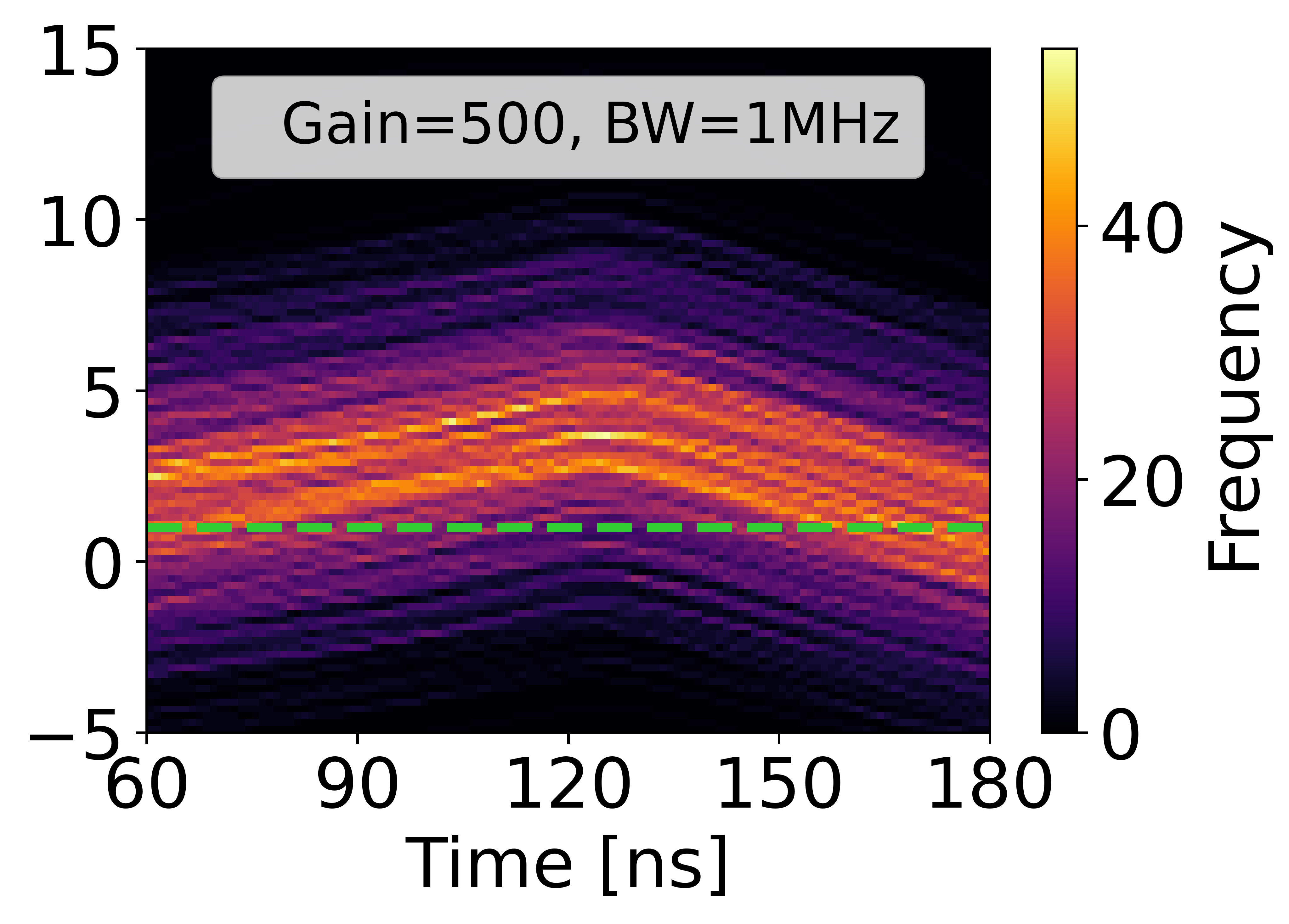}} 
    \hfill   
\subfloat[\label{fig: G1_BW120}]{
    \includegraphics[width=0.34\linewidth]{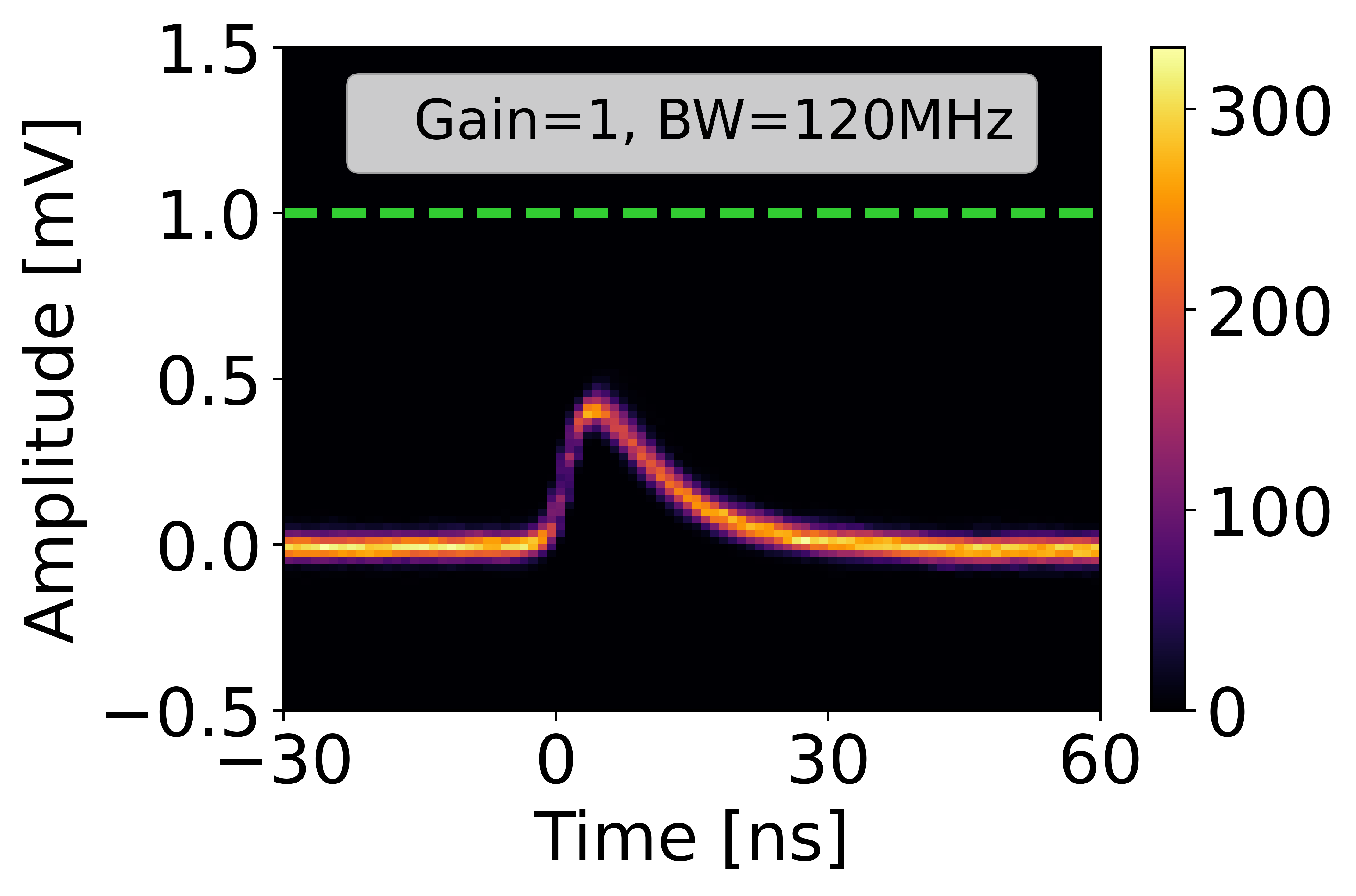}}
    \hfill 
\subfloat[\label{fig: G20_BW6}]{
    \includegraphics[width=0.30\linewidth]{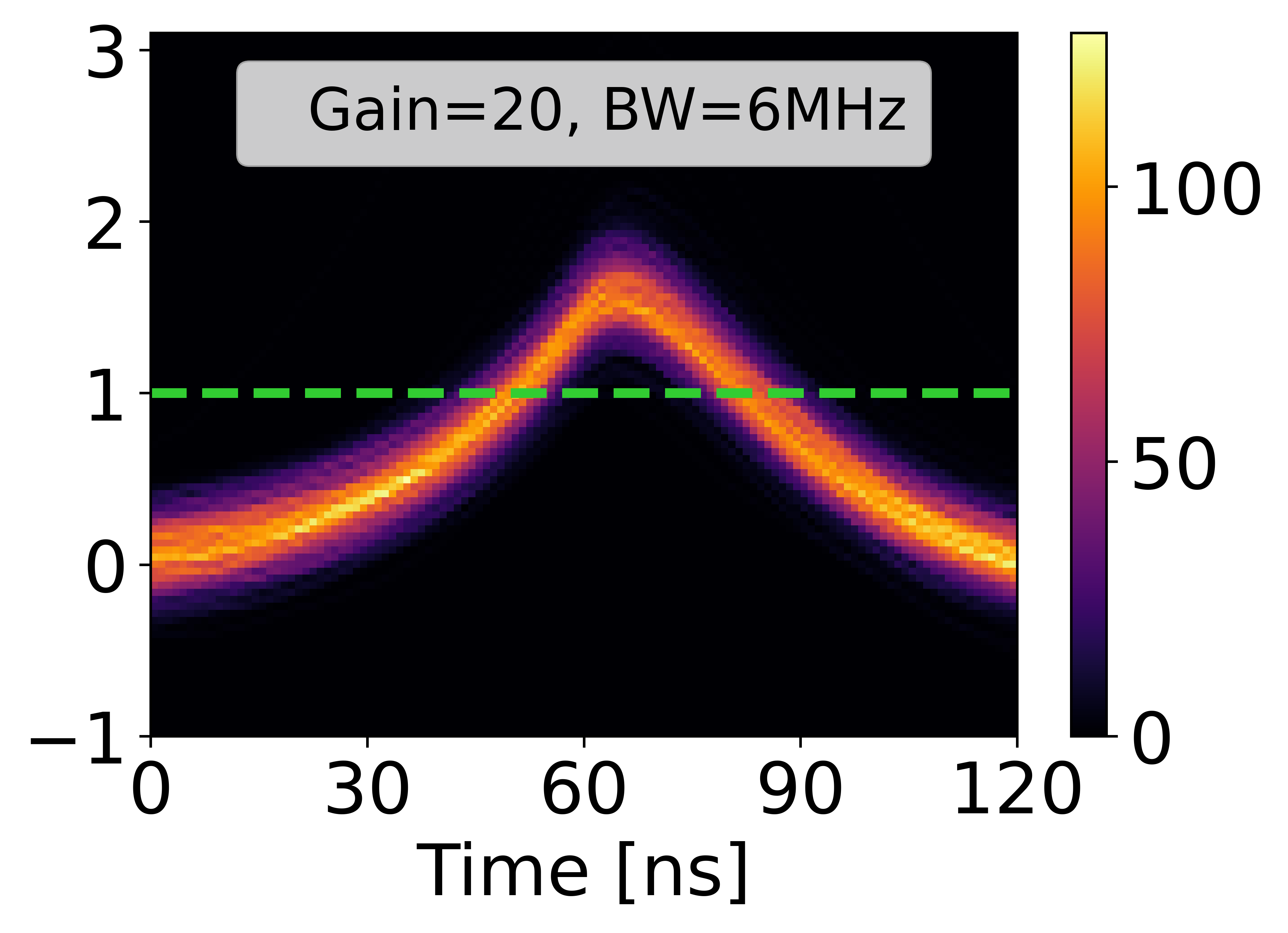}} 
    \hfill 
\subfloat[\label{fig: G120_BW1}]{
    \includegraphics[width=0.31\linewidth]{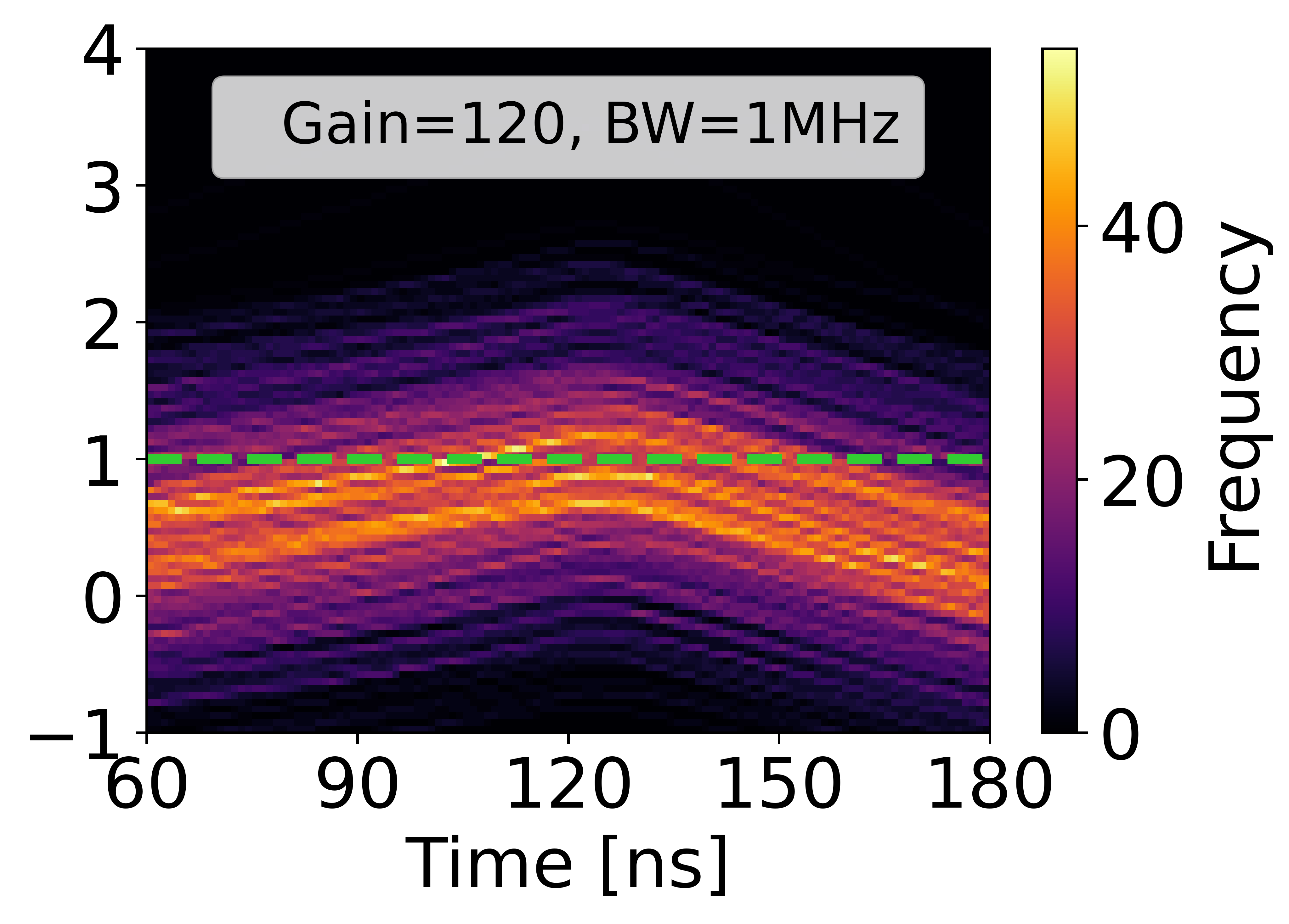}} 
    \hfill 
\subfloat[\label{fig: G1_BW80}]{
    \includegraphics[width=0.34\linewidth]{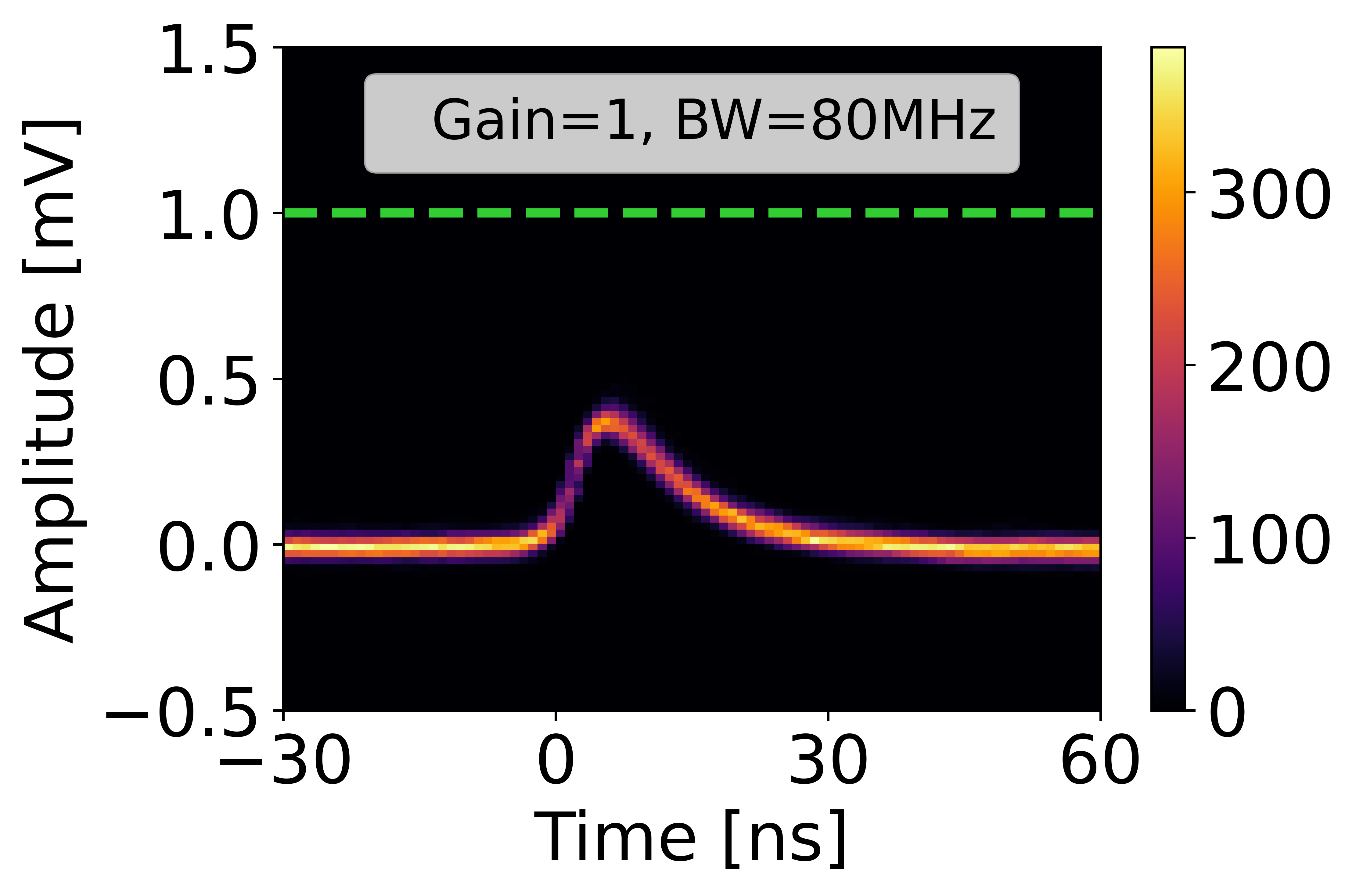}}
    \hfill 
\subfloat[\label{fig: G40_BW2}]{
    \includegraphics[width=0.30\linewidth]{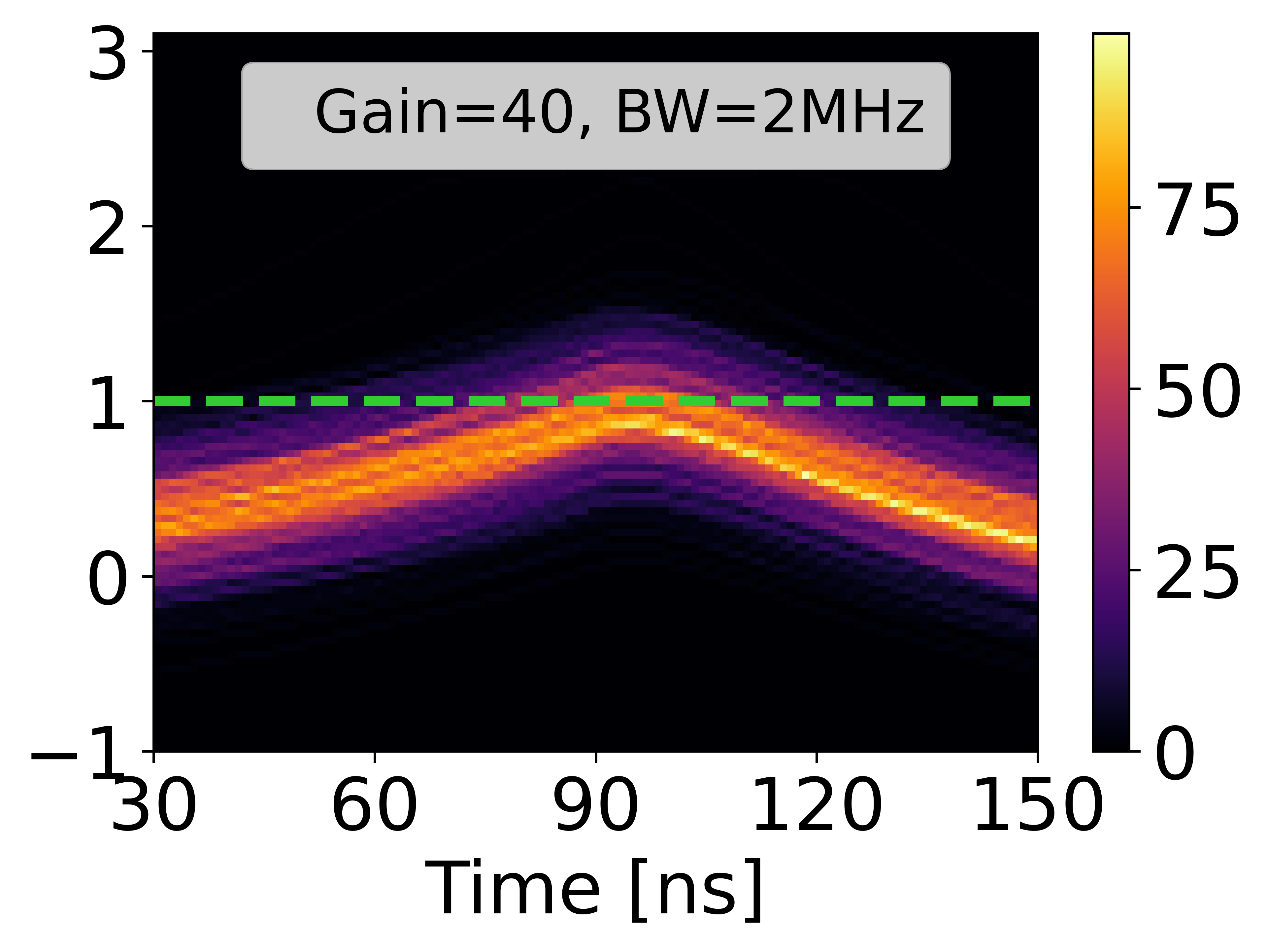}} 
    \hfill 
\subfloat[\label{fig: G80_BW1}]{
    \includegraphics[width=0.31\linewidth]{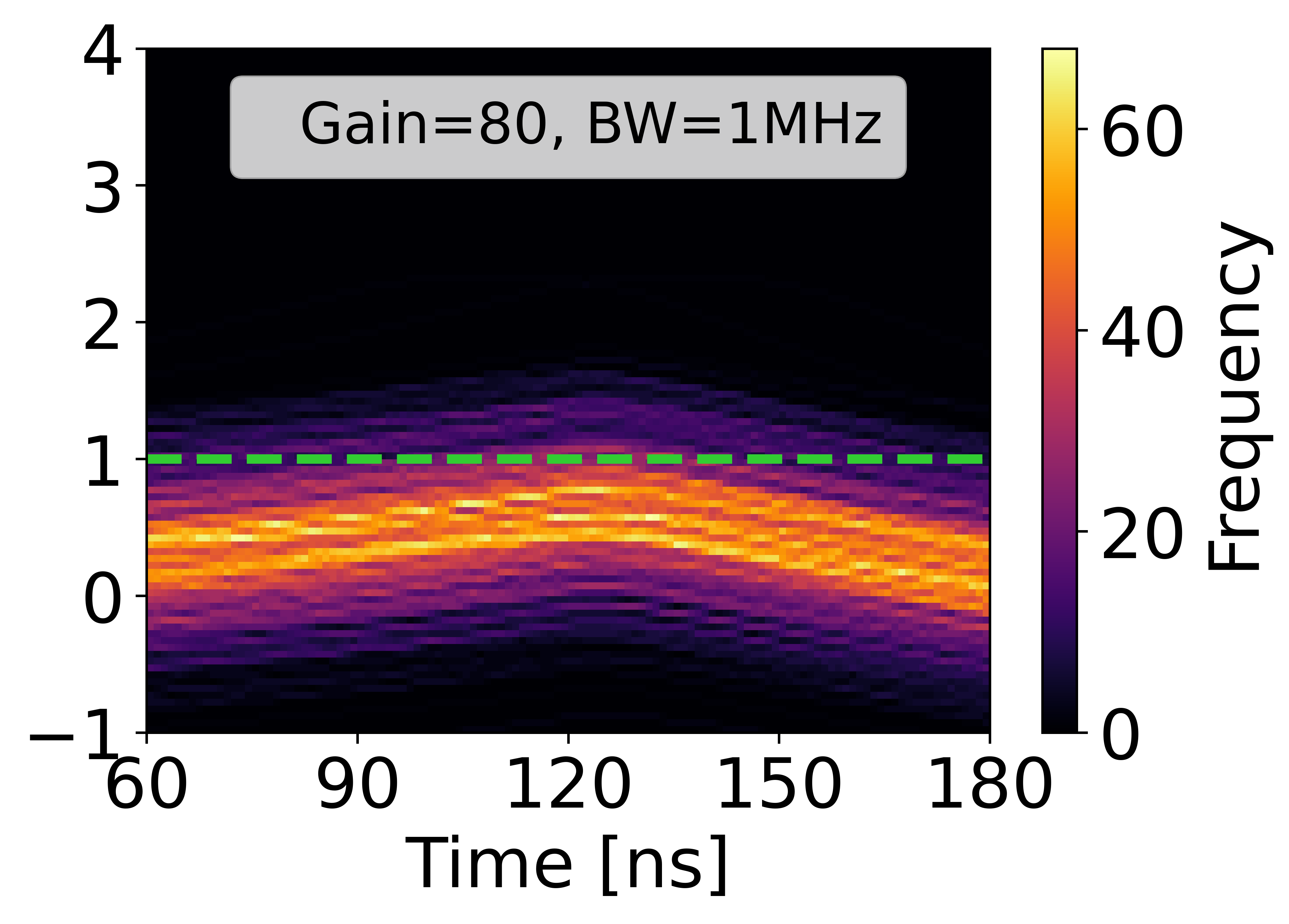}} 
    \hfill
\caption{The statistical SiPM standard output of 1000 pulse for different gain-bandwidth value combinations under the selected GBP. The green line shows the minimum comparator threshold $V_{th}$, which is 1~mV. (a) Gain=1, BW=500~MHz (b) Gain=10, BW=50~MHz (c) Gain=500, BW=1~MHz (d) Gain=1, BW=120~MHz (e) Gain=20, BW=6~MHz (f) Gain=120, BW=1~MHz (g) Gain=1, BW=80~MHz (h) Gain=40, BW=2~MHz.(I) Gain=80, BW=1~MHz}
  \label{fig: statistics under different GBW}
\end{figure*}

Based on the selected threshold, we chose the data rate of 100~kbps for the BER evaluation first. The BER performance was investigated by choosing six optical power intensities to cover a range of BER between $10^{-1}$ to $10^{-6}$. Fig.~\ref{fig: BER result}\subref{fig: FPGA threshold sweeping} shows the bathtub curves of the digital threshold $n_{T}$ for each bit versus the BER result, which is expected as (\ref{eq: PE}). The best BER performance was then chosen to plot for each optical power. Based on this step, the BER performance of the data rate from 10~kbps to 1~Mbps was investigated in Fig.~\ref{fig: BER result}\subref{fig: BER vs. optical power}. The maximum data rate of 1~Mbps under $2.45 \times 10^{-3}$ BER was achieved at -74.98~dBm with an average of 7.89 detected photons per bit, which has a 2.45\% difference from the Poisson limit for detected photons.

When the optical power level is below -70~dBm, the detected photon count is determined by the incident optical power. At a level of -70~dBm, the SiPM reaches its maximum count rate, enabling it to achieve a data rate of 1~Mbps. However, if the incident optical power exceeds -70~dBm, as illustrated in Fig.~\ref{fig: realtime result}\subref{fig: DR_offline_realtime}, the pulses begin to overlap, resulting in missed photon counts, leading to poor BER performance. Since the bit time for a data rate of 1~Mbps is 1~$\mu$s, which is approximately 30 times larger than the SiPM's recovery time, therefore the chance of non-linearity for the ISI is negligible compared with the pulse overlap at increased incident optical power.

The outer figure in Fig.~\ref{fig: data rate vs. counts per bit} shows the detected photon count per bit $\lambda_{s}$ penalty at PE of $10^{-3}$ from equation (\ref{eq: PE}). From the figure, the PE is computed for a receiver with an adaptive decision threshold, and a PE of $10^{-3}$ can be achieved by an average of 6.2 detected photons if there is no detected background photon. It is noticeable that when the background noise increases, the average signal detected photon count needed to sustain the BER also increases, resulting in an increase in power penalty. The inner figure shows the optical power required for different dark count rates under the data rate below 1 Mbps. The optical power is calculated based on the detected photon count per bit in the outer figure and the PDE at 620 nm of C series SiPM. The dark count rate is calculated as dark count per second. If the dark count rate is decreased while keeping the data rate constant, the number of dark counts per bit time is reduced. Consequently, it leads to a lower power penalty required to maintain a PE of $10^{-3}$ until the optical power reaches 6.2 detected photons per bit time. However, if the data rate is decreased while keeping the dark count rate constant, the receiver becomes more sensitive, but the power penalty increases as the data rate falls because of higher $\lambda_{b}$.

\subsection{Power Consumption}

To optimize the real-world performance of the real-time SiPM-based receiver for IoT applications, the power consumption of its components was measured. Table II presents the power consumption measurements for the prototyped receiver under a data rate of up to 1~Mbps. It is observed that the SiPM's power consumption increases with an increase in data rate. This is because the current within the SiPM originates from electrons excited by the detected photons, maintaining a proportionate relationship with the incident light. The ability to achieve a higher data rate depends on detecting more photons. In the meantime, the measured power consumption of the evaluation board was considerably higher than that of the designed circuit due to numerous unused peripheral interfaces, advanced reduced instruction set computer machine (ARM) core, and FPGA sources during the board's power-up process. To evaluate the power consumption of the designed receiver circuit, separate measurements were taken for the Xilinx ZYNQ 7000 FPGA, first with only the transmitter PRBS generator and then with both the transmitter and receiver implemented. The difference in these values gives an estimate of the power consumption of the digital circuit of the receiver, which is 36~mW. Among the receiver components, the three amplifiers consume the highest amount of power, which is approximately 2~W. Therefore, analyzing the power consumption of the amplifiers should be a focus in sections V and VI.

\section{Effect of Gain Bandwidth Product}

The previous section designed the receiver based on the ideal setup to investigate the SiPM performance. However, the receiver components often contain amplifier blocks and lowpass or bandpass hardware filters, which affect the shape of the SiPM output pulses to the FPGA. To ensure the best transmission performance of the SiPM pulses, three high GBP amplifier blocks were used in the real-time experiments. However, these high-performance amplifiers also increase the receiver's power consumption, a disadvantage, especially in IoT applications. When an amplifier is selected, the factors such as bandwidth, slew rate and power consumption should be considered. For a single-pole response voltage feedback amplifier, the product of the DC gain and the bandwidth is constant, which has a trade-off with power consumption \cite{Carter}. In order to minimize the power consumption of the receiver, the effect of the receiver's  GBP on the BER was investigated. Since changing the GBP of each amplifier is not practical due to experimental limitations, the rest of the investigation uses the numerical simulation based on the offline processing method in section II. The captured sample waveforms from the oscilloscope were filtered through a first-order Butterworth low pass filter (LPF) implemented in software with a bandwidth below 1~GHz. 

Fig.~\ref{fig: statistics under different GBW} shows the histograms of 1000 captured pulses after filtering within the various combinations between bandwidth and gain under a fixed GBP of 500~MHz, 120~MHz and 80~MHz. A minimum threshold $V_{th}$ of 1~mV (the green line) was chosen based on the input offset voltage of most commercial comparators from Texas Instruments (TI) and Analog Devices (ADI). The sweeping threshold method described in section II was also applied to obtain the detected photon count event for each GBP configuration. The hysteresis of 25\% $V_{th}$ was deployed to avoid the electrical noise impact. However, the hysteresis value could potentially be decreased due to limitations in bandwidth. With a lower bandwidth, the noise level experiences a decrease. In the offline measurements using an oscilloscope, a 5~mV hysteresis voltage was added based on the amplitude of DC noise shown in Fig.~\ref{fig: fast and standard pulse}\subref{fig: standard pulse}. However, in a real-time system, the appropriate hysteresis setting may vary depending on the noise spectrum in the experimental system, hence the total noise which impacts the hysteresis decisions.

Since the GBP in the experimental setup was significantly higher than required, the numerical simulation started from half of the experimental bandwidth, which is 500~MHz. As Fig.~\ref{fig: statistics under different GBW}\subref{fig: G1_BW500}\subref{fig: G10_BW50}\subref{fig: G500_BW1} show, increasing amplifier gain leads to a higher output pulse amplitude, whilst decreasing amplifier bandwidth causing longer pulse width, thus longer recovery times. A similar trend is found in the other two selected GBP simulation results in Fig.~\ref{fig: statistics under different GBW}\subref{fig: G1_BW120}\subref{fig: G20_BW6}\subref{fig: G120_BW1} and Fig.~\ref{fig: statistics under different GBW}\subref{fig: G1_BW80}\subref{fig: G40_BW2}\subref{fig: G80_BW1}. When the minimum $V_{th}$ of 1 mV was set, the statistical pulse amplitude in Fig.~\ref{fig: statistics under different GBW}\subref{fig: G1_BW500}\subref{fig: G1_BW120}\subref{fig: G1_BW80} could not reach $V_{th}$ due to the relatively low gain configuration, therefore no count events were recorded. By increasing the gain, Fig.~\ref{fig: statistics under different GBW}\subref{fig: G40_BW2}\subref{fig: G80_BW1} approach the $V_{th}$, but the majority counts are lost as most pulses are still below $V_{th}$. In Fig.~\ref{fig: statistics under different GBW}\subref{fig: G500_BW1}\subref{fig: G120_BW1}, the amplitude of the filtered pulses is higher than $V_{th}$ but noisy. It was observed that the very low bandwidths reduce the amplitude hence an appropriate gain and bandwidth configuration is required. However, if gain and bandwidth are more than the required minimum, then as shown in Fig.~\ref{fig: statistics under different GBW}\subref{fig: G20_BW6}\subref{fig: G10_BW50} there is a significantly higher noise margin, as the amplitude is well above the $V_{th}$. Overall, the best case of amplified SiPM pulse above the minimum $V_{th}$ is shown in the middle of each GBP group. 

\begin{figure}[t]
    \centering
    \includegraphics[width=3.4in]{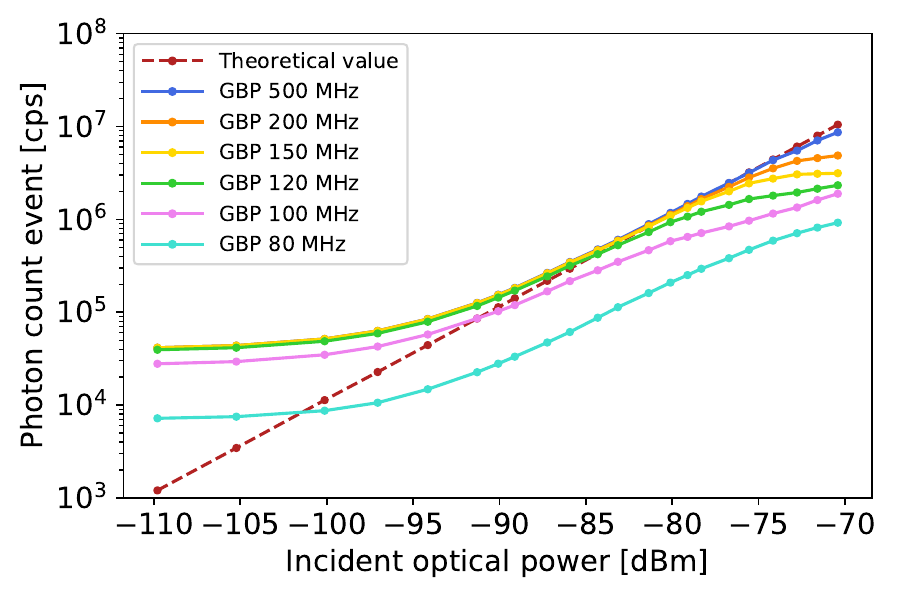}
    \caption{The offline processed detected photon count event as a function of incident optical power compared with the expected theoretical linear response under the various GBP from 80~MHz to 500~MHz.}
    \label{fig: DR under different BW}
\end{figure}

Fig.~\ref{fig: DR under different BW} shows the incident optical power versus count under different GBPs. The maximum simulated optical power was selected by the non-linear region that was found in Fig.~\ref{fig: offline result}\subref{fig: dynamic range}. In Fig.~\ref{fig: DR under different BW}, when the GBP was decreased from 500~MHz to 120~MHz, a deviation trend emerges between the count event and the theoretical limits. This deviation trend shows the effective dynamic range is reduced due to the increasing pulse width between Fig.~\ref{fig: statistics under different GBW}\subref{fig: G10_BW50} and Fig.~\ref{fig: statistics under different GBW}\subref{fig: G20_BW6}. In addition, the minimum $V_{th}$ of the comparator is unable to count events when the GBP is lower than 120~MHz, which was explained in Fig.~\ref{fig: statistics under different GBW}\subref{fig: G40_BW2}. Overall, the minimum GBP to maintain the photon counting ability at the receiver is approximately 120~MHz.   

\begin{figure}[t]
    \centering
    \includegraphics[width=3.4in]{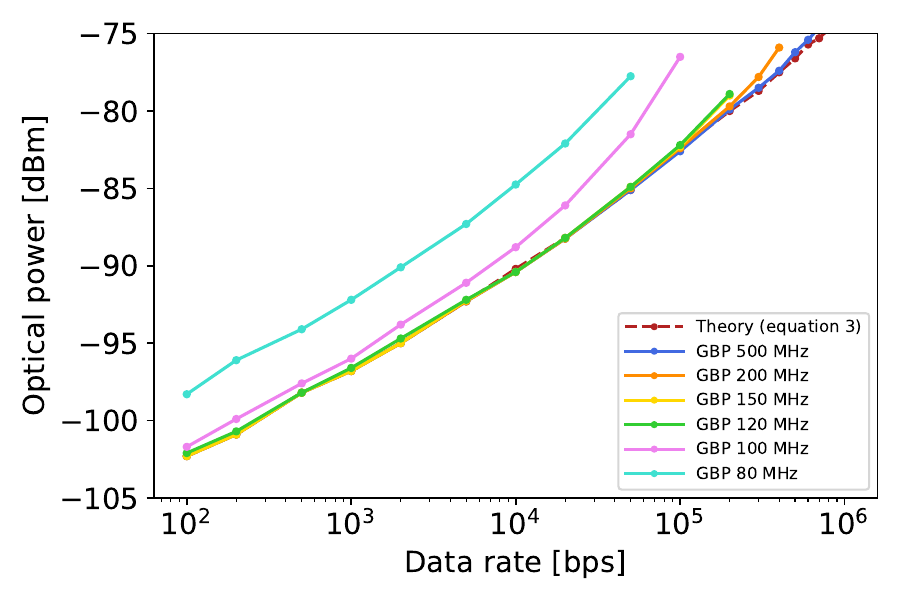}
    \caption{Optical power versus data rate at fixed target BER of $10^{-3}$ for the various GBP.}
    \label{fig: power vs data rate vs BW}
\end{figure}

Fig.~\ref{fig: power vs data rate vs BW} presents the simulation result between data rate versus incident optical power at BER of $1 \times 10^{-3}$ under different GBP. The BER was calculated based on the detected photon events per bit through (\ref{eq: PE}). The theoretical Poisson limit value calculated from (3) to achieve the maximum data rate of 1~Mbps at a PE of $1 \times 10^{-3}$ is approximate -74~dBm, with 9.32 detected photons per bit when the $\lambda_{b}$ is 0.048. It is observed that the 120~MHz is the minimum GBP to maintain the theory BER of $1 \times 10^{-3}$ and is the same as the minimum GBP that maintains the photon counting ability. In addition, a higher target data rate requires higher GBP, because of the increased maximum count rate when the optical power is higher. However, if the SiPM is saturated, the measured BER will deviate from the theory PE, which was calculated from (\ref{eq: PE}). 

\begin{figure}[t]
    \centering
    \includegraphics[width=3.4in]{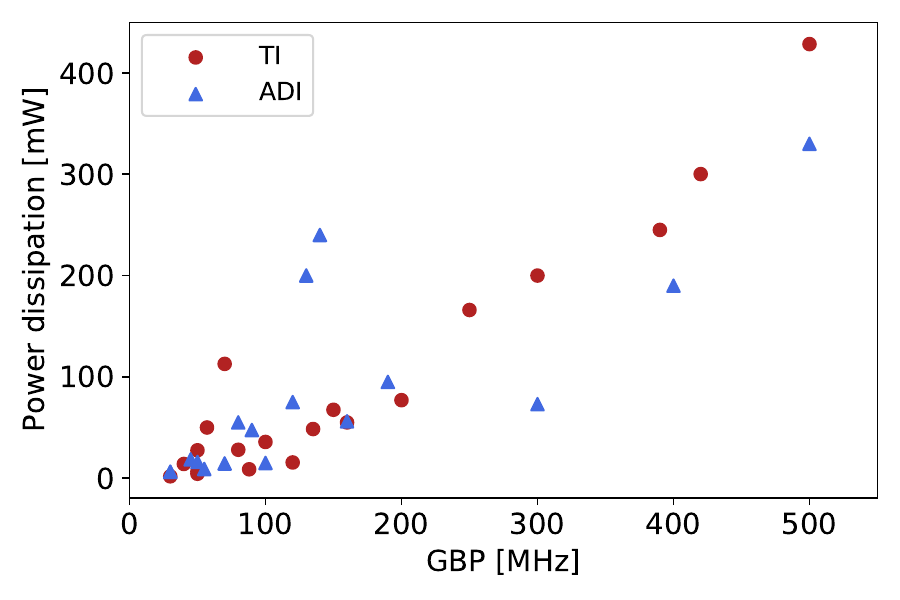}
    \caption{Electrical power dissipation versus GBP with typical commercial Op-amps.}
    \label{fig: Power consumption vs GBW}
\end{figure}


\section{Discussions}
\subsection{Further Improvements and Contributions}

Considering the bandwidth limitation of the PMOD connector on the FPGA evaluation board, we focused on the SiPM standard output to present the SiPM's dynamic range and Poisson limited BER performance, as well as the bandwidth limitation on the SiPM readout circuit. We have experimentally verified that the Poisson limit can be approached at the data rate between 10~kbps to 1~Mbps when the optical power was below -74~dBm for the standard output. However, the high sensitivity of the SiPM was achieved by counting individual pulses in each bit period, which needs a high GBP analog readout circuit, hence, a higher power consumption receiver circuit is required\cite{Carter}. In order to practically demonstrate this concept, the power consumption of 35 commercially available op-amps from TI and ADI was calculated from their respective data sheets and shown in Fig.~\ref{fig: Power consumption vs GBW}. According to the simulation results in Fig.~\ref{fig: DR under different BW}, a minimum GBP of 120~MHz is needed to preserve the photon counting capability while maintaining a power consumption of 50~mW shown in Fig.~\ref{fig: Power consumption vs GBW}. This 50~mW value represents a significant reduction in power consumption compared to the prototype amplifiers used in real-time setup.

The high GBP requirement of the readout circuit is also similar to the other photon counting detectors, including photomultiplier tubes (PMT) \cite{Ning} and  other modulation schemes such as pulse position modulation (PPM) \cite {popoola} which is used in many applications, such as VLC\cite{rios} and UWOC\cite{liu} due to its high power efficiency and noise immunity. Based on the simulation result, the minimum required GBP to maintain the photon counting ability is 120~MHz. 

The obtained results for the standard output of SiPM were not optimum due to the larger single-photon pulse width compared with the fast output, leading to a smaller dynamic range. However, we can estimate that the fast output or other SiPMs featuring shorter pulse widths increase the maximum count rates, ultimately leading to higher data rates beyond 1~Mbps. As the use of other series SiPMs with shorter pulse widths will perform a larger dynamic range and maximum count rate, ultimately leading to higher data rates. In the future, to achieve a higher dynamic range and data rates, a higher bandwidth interface on the FPGA evaluation board is required, for example, FPGA mezzanine card (FMC). However, the implementation of fast output still requires higher GBP amplifiers which increase the electrical power consumption even further, and higher bandwidth FMC connectors increase the overall complexity and cost, especially for IoT applications. 

Moreover, since the current commercially available SiPMs have a higher PDE in the visible blue-green spectrum, for example, in UWOC, VLC and Li-Fi applications, it is expected that a lower optical power is required to achieve the same BER at a longer wavelength. However, these SiPM are not yet suitable for near-infrared (NIR) communication, such as infrared data association (IrDA), which is typically 850~nm. The NIR SiPMs are expected in the near future due to progress in SPAD fabrication technology \cite{van2022backside, Saha}. The theory and experimental results demonstrated in this work remain valid for any wavelength range since it is based on the detected photons.

\begin{table*}[t]
\centering
\renewcommand{\arraystretch}{1.2}
\caption{Comparative Analysis of Transmitter Power Consumption: The Proposed System vs. Bluetooth Modules}
\label{tab: Tx Power Consumption}

\begin{tabular}{ccccc}
\toprule
\toprule

\multicolumn{2}{c}{Tx module name}                     & Data rate                                                                    & Voltage      & Tx active current      \\ \Cline{1-5}{1pt}
\multicolumn{2}{c}{AIROC™ CYW20822-P4TAI040}          & \begin{tabular}[c]{@{}c@{}}125 kbps, 500 kbps,\\  1 Mbps, 2 Mbps\end{tabular} & 3.3V         & 1.4 mA to 4 mA         \\ \Cline{1-5}{1pt}
\multicolumn{2}{c}{SmartBond™ DA14531}                & 1 Mbps                                                                       & 3V           & 1.5 mA to 5mA           \\ \Cline{1-5}{1pt}
\multicolumn{2}{c}{Laird Connectivity BL651}          & 1 Mbps, 2 Mbps                                                               & 3V           & 2.1 mA to 7 mA          \\ \Cline{1-5}{1pt}
\multirow{2}{*}{MCU + LED} & C8051F98x MCU            & Adjustable                                                                   & 3.3 V        & 150 $\mu$A/MHz               \\ \Cline{2-5}{0.01pt} 
                           & HLMP-EG1A-Z10xx LED      & 10 kbps to 1 Mbps                                                                 & 1.6 V to 2 V & 0.3 mA to 10 mA      \\

\toprule
\toprule
\end{tabular}
\end{table*}

In comparison with existing research, below are the main contributions of this work.

1)	To the best of our knowledge, this design represents the first real-time photon counting receiver implementation on a conventional SiPM and an FPGA, enhancing its potential for IoT applications compared to previous offline approaches \cite{Zubair1},\cite{Zubair2}, \cite{Shenjie}, \cite{Jinjia}, \cite{Long1}.

2)	By conducting numerical simulations, this study assessed the GBP of the post-readout circuit within the SiPM-based optical receiver. This assessment complements previous research findings and offers insights into the circuit's suitability for future low-power consumption applications.

3)	An FPGA-based customized design is implemented for photon accumulation. This approach uses a dual-interleaved counter to ensure uninterrupted photon counting during each bit, avoiding the dead time compared to the one sequential counter \cite{sampad2021fpga}. Additionally, utilizing an asynchronous detection mechanism for the counters eliminates the need for a high-frequency sampling clock, simplifying the design for low-data rate systems.

4)	The proposed SiPM-based receiver approaches the Poisson limit in photon detection, achieving a BER of $2.45 \times 10^{-3}$ with 7.89 detected photons per bit compared to 7.7 detected photons based on equation (3). Regarding the incident photons distance to the Poisson limit, these detected photons indicate approximately 219 incident photons, a 14.4 dB gap based on a PDE of 3.6\%. Compared with the sensitivity of other photodetectors, although the APD is proven more sensitive than PIN PD \cite{kharraz2013performance,promise2018sensitivity}, their sensitivities are a few orders away from the Poisson limit. In this situation, the SiPM is potentially more sensitive than the APD if the SiPM operates at data rates where the bit time is longer than the output pulse width \cite{zhang2018comparison}. Further improvements in avalanche diode fabrication technology similar to \cite{van2022backside} are expected to bring the incidents closer to the Poisson limit because of improvements in the PDE. 

\subsection{Potential Scenarios for Application Use}

In numerous IoT applications, such as smart homes, wearable devices, and monitoring systems, there has been a traditional dependence on radio frequency (RF) technology. Typically, in RF-based systems, sensor data is transmitted using Bluetooth Low Energy (BLE), ideal for peer-to-peer (P2P) connections requiring low data transmission rates over short distances \cite{afaneh2018intro}. When designing a BLE transmitter, it is essential to consider several factors, such as the antenna design, signal strength, and the modulation of the 2.4 GHz baseband. 

Compared to the BLE, the proposed SiPM-based receiver enables an alternative approach in which the low-power sensors use LEDs to transmit slow-changing data like air quality, temperature, and pressure. This approach is efficient for environmental, medical, and industrial monitoring. Considering the requirement of low power consumption and long-term functionality for multiple sensor transmitters, the design of the sensor transmitter board can incorporate a low-power microcontroller unit (MCU) to collect sensor data and transmit it via an LED, which can be designed to connect to a general purpose input/output (GPIO) port and include a resistor to limit the current. Power-saving strategies can also be applied within the MCU to switch the transmitter to a low-power mode when it is not actively used. Since the SiPMs-based receivers may require more power than standard IoT devices, integrating them into IoT hubs is a strategic choice that prioritizes enhanced optical sensitivity over electrical power efficiency concerns. 

In the experimental setup illustrated in Fig.~\ref{fig: offline setup} and Fig.~\ref{fig: realtime setup}, considering the LED operating near its forward voltage and the receiver requiring only low light intensity, an ND20 Filter is used to attenuate the light intensity of the LED. The reduced light intensity enables the SiPM to operate in its linear response range and enhance the measurement accuracy. 

Table III shows that the transmitter power consumption of commercially available BLE modules is higher than that of an LED-based transmitter. For example, with the MCU running at 1 MHz, the MCU power usage is at 495 $\mu$W. Considering the 100 times intensity attenuation due to the ND20 filter, an appropriately sized LED's current consumption is approximately 100 $\mu$A. Hence, the total power consumption needed for LED transmitters is around 695 $\mu$W, significantly less than the power consumption of BLE modules, which ranges from several to tens of milliwatts. However, because the LED operates near its forward voltage, the effective current consumption of the LED might be less than 100 times the attenuated value by the ND20 filter. Also, it is essential to note that large LEDs may not achieve micro-watt level power consumption. Conversely, smaller LED dies can achieve lower power consumption, allowing for power usage at the micro-watt level\cite{cho2020monolithic, huang2020mini}. 

Additionally, the reduced power consumption of LED-based transmitter benefits from the LED focused light transmission, in contrast to the omnidirectional transmission of radio waves antenna in BLE systems, resulting in more significant power usage. Moreover, with a 2.4 GHz baseband, the trace antenna's length is approximately 31 mm. This dimension is larger than the typical standard 100  $\mu$m square LED die, resulting in more board space occupied.
 

\section{Conclusions}
In this paper, we have demonstrated a novel real-time SiPM-based receiver with a low bit rate and high sensitivity, which has the potential for low transmitter power consumption. The work provides the evaluations of the analog chain of the receiver to show the potential for lower power consumption. The numerical simulation proves that the required power consumption of the amplifier is approximately 50~mW at 120~MHz GBP. In addition, to further reduce the complexity and power consumption in the digital circuit design, the FPGA implemented an asynchronous photon detection method. Finally, the implementation of interleaved counters in the receiver allows it to receive streaming data without dead time. This design is being implemented on an FPGA and conventional SiPM for the first time to the best of our knowledge, making it more beneficial for utilizing SiPM in IoT applications than previous offline approaches.

During the characterizations of SiPM's standard output, the average FWHM of pulses was approximately 8~ns. A voltage thresholding method was developed and implemented using an op-amp-based comparator with hysteresis to count the detected photons avoiding the background electrical noise. The final configured threshold was decided based on sweeping this threshold voltage in numerical simulation and real-time hardware. Results show that both offline and real-time voltage threshold methods approached the theoretical Poisson limit through the standard output of the SiPM. In addition, the real-time system demonstrated a BER of $2.45 \times 10^{-3}$ at a data rate of 1~Mbps under an incident optical power of \mbox{-74.98}~dBm, using a 620~nm LED. This result was achieved with an average of 7.89 detected photons per bit, which has a 2.45\% difference considering the Poisson limit for detected photons. The detected photons suggest an estimated total of 219 incident photons, indicating a 14.4~dB gap from the Poisson limit for incident photons, considering a PDE of 3.6\%. While this gap might be reduced by using a SiPM with higher PDE, the difference in the Poisson limit for detected photons will remain the same regardless of the choice in SiPM. Moreover, the relationship between the GBP of the SiPM readout circuit and the target data rate to achieve a BER of $10^{-3}$  was evaluated through numerical simulations and offline data processing. The simulation results show that receiver GBP needs at least 120~MHz with a bandwidth of 6~MHz to maintain the counting ability when the comparator threshold is set to 1~mV. The reason behind the limitation in the SiPM's counting ability is the bandwidth restriction below 6 MHz, which leads to a significant increase in the pulse width. Overall, to maintain photon-counting ability at the receiver, it is necessary to have a minimum GBP of around 120 MHz when the gain and bandwidth settings are configured to 20 and 6 MHz, respectively.

To maintain Poisson-limited photon counting capability, the necessary bandwidth is considerably higher than the target data rate. In comparison to detectors like photodiodes, bandwidth is approximately equal to 65\% of the data rate in \mbox{OOK} non-return-to-zero (NRZ) without achieving photon counting sensitivity \cite{Eduard}. This suggests that attaining enhanced sensitivity nearing the Poisson limit leads to an increase in power consumption. Such a receiver has the potential to be utilized in central IoT hubs where electrical power consumption is not a major priority. However, high optical sensitivity is required due limited energy of the optical transmitter on IoT client devices, which transmit sensory information at low data rates. In the future, further investigation is needed to implement and assess the proposed use case scenarios.

\section*{Acknowledgment}

The authors would like to thank Prof. Robert Henderson for providing lab space and equipment for this work, and Dr Shenjie Huang and Prof. Majid Safari for their influential discussions during this work.

\ifCLASSOPTIONcaptionsoff
  \newpage
\fi

\bibliographystyle{IEEEtran}
\bibliography{reference}

\end{document}